\documentclass[11pt]{article}
\usepackage[letterpaper, left=1in, top=1in, right=1in, bottom=1in,nohead,includefoot]{geometry}
\usepackage{color,charter,graphicx,here,setspace,multirow,amsmath,amssymb,amsfonts,enumitem,placeins,amsthm,cancel,mathrsfs,bm,pdflscape,subcaption}
\usepackage[utf8]{inputenc}
\usepackage[authoryear]{natbib}
\usepackage{ulem} 

\usepackage[dvipsnames]{xcolor}
	\usepackage[pdftex]{hyperref}
	\hypersetup{colorlinks,
	citecolor=NavyBlue  
	}

\usepackage{subfiles}

\usepackage[colorinlistoftodos,prependcaption,textsize=tiny]{todonotes}

\usepackage{mathrsfs}
\usepackage{amssymb}
\usepackage{amsmath}
\usepackage{mathtools}
\usepackage{ascmac}
\usepackage{amsthm}
\usepackage{multirow}
\usepackage{proba}
\usepackage{subcaption}
\usepackage{adjustbox}

\usepackage{graphicx}
\usepackage{natbib}
\usepackage{setspace}
\usepackage{bm}
\usepackage{url}

\usepackage{color}

\usepackage{times}

\newtheorem{thm}{Theorem}

\begin{document}

\title{Optimal Hold-Out Size in Cross-Validation}
\author{Kenichiro McAlinn\thanks{Department of Statistical Science, Fox School of Business, Temple
University, Philadelphia, PA. {\scriptsize{}Email:
kenichiro.mcalinn@temple.edu}}\, \& K\={o}saku Takanashi\thanks{RIKEN Center for Advanced Intelligence Project {\scriptsize{}Email: kosaku.takanashi@riken.jp} }}

\maketitle
\thispagestyle{empty}\setcounter{page}{0}
\begin{abstract}
Cross-validation (CV) is routinely used across the sciences to select models and tune parameters, and the resulting choices are often interpreted as substantive scientific conclusions (e.g., which variables, mechanisms, or risk factors are ``supported by the data''). A key part of the CV procedure-- the hold-out size, or equivalently the fold count $K$-- is typically set by convention (e.g., 80/20, $K=5$) rather than by a principled criterion. Central to the issue is the tradeoff between training and testing: increasing the training sample size improves model accuracy, while sacrificing certainty around the accuracy itself. We formalize the tradeoff by targeting predictive performance and explicitly penalizing evaluation uncertainty, which cannot be identified from the data without additional assumptions. We derive finite-sample expressions of this evaluation uncertainty under symmetric errors and general upper bounds under broader error conditions, yielding a transparent utility-based rule for selecting the hold-out size as a function of an irreducible-noise parameter. Empirical analyses with linear regression and random forests across multiple domains, and a high-dimensional genomics application, show that (i) the choice of $K$ is dependent on the data and model. (ii) the optimal $K$ varies based on the assumption on the irreducible error, and (iii) the implied inferential conclusions can change materially as the irreducible error, and thus $K$, varies. The resulting framework replaces a one-size-fits-all convention with a context-specific, assumption-explicit choice of $K$, enabling more reliable model comparisons and downstream scientific inference.

\bigskip{}
\end{abstract}
\newpage{}

\section*{Introduction}
Cross-validation (CV) is widely treated as a ``neutral'' way to evaluate predictive performance and
choose among models or tuning parameters. In practice, however, CV is often used as a {gatekeeper for
scientific claims}: a model choice made by CV can determine which covariates are reported as significant,
which mechanisms are described as active, or which policies are recommended. In these settings, the CV
procedure is not merely a computational device; it is part of the inferential pipeline. A central and
rarely justified component of that pipeline is the hold-out size, or equivalently, the number of folds
$K$. Most applied work defaults to convention (80/20, $K=5$, $K=10$), even though changing $K$ can change
the selected model and the conclusions drawn from it.

The reason is simple: $K$ controls two competing effects. Larger training sets can yield better-fitted
models (lower predictive bias), but smaller test sets yield noisier performance estimates (higher
evaluation variance). When predictive performance is strongly separated across candidate procedures, these
differences may not matter. However, in many modern problems-- especially with flexible models, many tuning
parameters, or high-dimensional sparsity-- the performance curve can be nearly flat. In that regime, the
choice of $K$ can act primarily on {stability and selection} rather than on average fit: two choices of
$K$ may have essentially identical estimated loss, yet produce different selected models and, therefore,
different downstream narratives. A researcher faced with conflicting CV outcomes across fold counts has no
standard, principled way to decide which $K$ is appropriate.

This paper treats the choice of hold-out size as a decision problem. Our target is {pure} predictive
performance-- performance net of irreducible noise-- rather than raw empirical loss, and we make explicit
that any decision based on CV must confront uncertainty in the evaluation itself. The main obstacle is that
evaluation uncertainty in $K$-fold CV is not universally estimable from the observed data without further
assumptions \citep{bengio2004no}. Existing theory provides valuable asymptotic descriptions of CV-based
comparisons and tests \citep[e.g.,][]{austern2020asymptotics,bayle2020cross,li2023asymptotics}, and guidance
for particular settings \citep[e.g.,][]{arlot2016choice}, but does not yield a general finite-sample,
dataset-specific criterion that transparently trades off accuracy against evaluation uncertainty.

Our contribution is to provide such a criterion by deriving finite-sample characterizations of the
evaluation uncertainty (exact under symmetric errors; upper bounds more generally), linking this
uncertainty to the loss, and embedding the tradeoff in an explicit mean--variance utility. This yields a
transparent rule that maps an irreducible-noise parameter (reflecting the analyst's tolerance for
evaluation uncertainty) to an optimal hold-out size. In contrast to a single ``recommended'' $K$, the
output is a robustness map: which $K$ is optimal under which noise regimes, and how conclusions change
across that map.

A final practical point concerns implementation. The theoretical analysis concerns evaluation on a single
hold-out set of size $m$, representing the fundamental uncertainty of assessing a fitted model on $m$ new
observations. In standard $K$-fold CV, $m=N/K$ and the reported loss averages across folds; this averaging
reduces variance but does not remove the underlying tradeoff induced by the hold-out size. Thus, choosing
$K$ remains a scientific decision: it determines how aggressively one prioritizes training accuracy versus
reliable evaluation.

The remainder of the paper formalizes this tradeoff, derives finite-sample results that make its
components explicit, and demonstrates the implications on real data. In particular, we show that the
optimal hold-out size depends on both the model class and the dataset, and that conventional choices
implicitly encode assumptions about irreducible noise and acceptable evaluation uncertainty. By making
these assumptions explicit, the proposed framework replaces convention with a principled, transparent
choice of $K$ and clarifies how that choice propagates to downstream inference.

\section*{Definitions and main theoretical result}
To derive our central theoretical results, we first define the notation, data-generating assumptions, and explicate cross-validation, including its key theoretical properties.
For the data generating assumptions, we consider cases where the irreducible error is symmetric (e.g., Gaussian), and where it is asymmetric (e.g., Gamma).
Given the definitions, we derive the theoretical exact or upper bound of the predictive accuracy uncertainty for both assumptions.

Denote the training set sample size as $n$ and the test set sample size
as $m$. The total sample size is $n+m=N$. 
Given a fixed total sample
size, an increase in $n$ means a decrease in $m$, and vice versa.
The central question is determining $m$ that best balances model accuracy and evaluation certainty, which in turn implies $K$.

For $K$-fold cross-validation, the test set size is $m=N/K$, where
$N$ is divided into $\{C_{1},...,C_{k},...,C_{K}\}$ sets. The test
set is $C_{k}$ and the training set is denoted as $C_{-k}=\{C_{1},...,C_{k-1},C_{k+1},...,C_{K}\}$
to represent the divided dataset minus the test set, $C_{k}$. The
training set consists of $\left\{ \left(Y_{i},X_{i}\right)\right\} _{i=1}^{n}$
and the test set consists of $\left\{ \left(Y_{j},X_{j}\right)\right\} _{j=n+1}^{n+m}$.
Throughout, we assume that the response follows the data-generating
process 
\[
Y_{i}=f(X_{i})+\varepsilon_{i},
\]
where the outcome is generated from some function of covariates, $f(X_{i})$,
and consider cases where $\varepsilon_{i}$ is symmetric and where it is asymmetric.
Together, this covers most problems encountered in data analysis.

For notational simplicity, we denote data and predictions on the test
set as $C_{k}$ and the training set as $C_{-k}$. We consider the squared error loss, 
\begin{alignat*}{1}
\hat{L}_{k}\left(\hat{\boldsymbol{\mu}}_{k}\right) & =\frac{1}{m}\left\Vert \hat{\boldsymbol{\mu}}_{k}-\boldsymbol{y}_{k}\right\Vert _{\mathbb{R}^{m}}^{2}-\frac{1}{m}\left\Vert \boldsymbol{\varepsilon}\right\Vert _{\mathbb{R}^{m}}^{2}\\
 & =\frac{1}{m}\sum_{j\in C_{k}}\left\{ \left(\hat{{f}}_{{-k}}\left(x_{j}\right)-y_{j}\right)^{2}-\varepsilon_{j}^{2}\right\} ,
\end{alignat*}
where $\hat{\boldsymbol{\mu}}_{k}=\left[\hat{f}_{{-k}}\left(x_{n+1}\right),...,\hat{f}_{{-k}}\left(x_{n+m}\right)\right]^{\top}$
is the prediction generated from the model trained on the training
set, $C_{-k}$, to predict the test set, $C_{k}$, and $\boldsymbol{y}_{k}=[y_{n+1},...,y_{n+m}]^{\top},\boldsymbol{\varepsilon}_{k}=\left[\varepsilon_{n+1},\cdots,\varepsilon_{n+m}\right]^{\top}$.
The $K$-fold cross-validation loss is 
\[
\hat{L}_{K\text{-CV}}\left(\hat{\boldsymbol{\mu}}_{k}\right)=\frac{1}{K}\sum_{k=1}^{K}\left\{ \frac{1}{m}\left\Vert \hat{\boldsymbol{\mu}}_{k}-\boldsymbol{y}_{k}\right\Vert _{\mathbb{R}^{m}}^{2}-\frac{1}{m}\left\Vert \boldsymbol{\varepsilon}_{k}\right\Vert _{\mathbb{R}^{m}}^{2}\right\} .
\]
In other words, these expressions quantify how close the model’s predictions are to the true data-generating process, averaged over all test sets.
The fundamental tradeoff in cross-validation is that as $n$ increases,
$\hat{f}_{-k}$ becomes more accurate (more learning), but the uncertainty
of $\hat{L}_{k}\left(\hat{\boldsymbol{\mu}}_{k}\right)$ increases,
due to a decrease in $m$.

Cross-validation is popular due to its ease of use (only requiring
data splitting, training, and predicting), but also due to its theoretical
properties, namely that it is an unbiased estimate of the prediction
loss (and a conditional unbiased estimate of pure loss): 
\begin{alignat}{1}
 & \mathbb{E}_{Y}\left[\left.\frac{1}{m}\left\Vert \hat{\boldsymbol{\mu}}_{k}-\boldsymbol{y}_{k}\right\Vert _{\mathbb{R}^{m}}^{2}-\frac{1}{m}\left\Vert \boldsymbol{\varepsilon}\right\Vert _{\mathbb{R}^{m}}^{2}\right|\boldsymbol{x}\right]=\nonumber \\
 & \phantom{\mathbb{E}_{Y}\left[\left.\frac{1}{m}\left\Vert \hat{\boldsymbol{\mu}}_{k}-\boldsymbol{y}_{k}\right\Vert _{\mathbb{R}^{m}}^{2}-\frac{1}{m}\right|\boldsymbol{x}\right]}\frac{1}{m}\left\Vert \hat{\boldsymbol{\mu}}_{k}-\boldsymbol{\mu}_{k}\right\Vert _{\mathbb{R}^{m}}^{2},\\
 & \mathbb{E}_{Y}\left[\left.\frac{1}{K}\sum_{k=1}^{K}\left\{ \frac{1}{m}\left\Vert \hat{\boldsymbol{\mu}}_{k}-\boldsymbol{y}_{k}\right\Vert _{\mathbb{R}^{m}}^{2}-\frac{1}{m}\left\Vert \boldsymbol{\varepsilon}_{k}\right\Vert _{\mathbb{R}^{m}}^{2}\right\} \right|\boldsymbol{X}\right]\thickapprox\nonumber \\
 & \phantom{\mathbb{E}_{Y}\left[\left.\frac{1}{K}\sum_{k=1}^{K}\left\{ \left\Vert \hat{\boldsymbol{\mu}}_{k}\right\Vert _{\mathbb{R}^{m}}^{2}\right\} \right|\boldsymbol{X}\right]}\frac{1}{mK}\left\Vert \hat{\boldsymbol{\mu}}_{k}-\boldsymbol{\mu}_{k}\right\Vert _{\mathbb{R}^{mK}}^{2},\label{eq:kCV}
\end{alignat}
for the basic cross-validation and $K$-fold cross-validation, respectively,
where $\boldsymbol{\mu}_{k}=[{f}(x_{n+1}),...,{f}(x_{n+m})]$ denotes
the vector of true response functions, and $\sigma^{2}=\mathbb{E}[\varepsilon^{2}]$
is the irreducible error. 
The approximation in \eqref{eq:kCV} is a notational simplification that treats the concatenation of fold-wise test vectors as an $mK$-vector; it is used only to motivate the shared loss-variance tradeoff and does not assume independence across folds.
While this loss can be empirically estimated,
except for the irreducible error, the variance of this loss, 
\begin{align*}
 & \mathbb{V}\textrm{ar}\left[\hat{L}_{m}\left(\hat{\mu}_{k}\right)\right]=\mathbb{E}_{Y}\left[\left(\frac{1}{m}\left\Vert \hat{\boldsymbol{\mu}}_{k}-\boldsymbol{y}_{k}\right\Vert _{\mathbb{R}^{m}}^{2}-\frac{1}{m}\left\Vert \boldsymbol{\varepsilon}\right\Vert _{\mathbb{R}^{m}}^{2}\vphantom{\mathbb{E}_{Y}\left[\left.\frac{1}{m}\left\Vert \hat{\boldsymbol{\mu}}_{k}-\boldsymbol{y}_{k}\right\Vert _{\mathbb{R}^{m}}^{2}-\frac{1}{m}\left\Vert \boldsymbol{\varepsilon}\right\Vert _{\mathbb{R}^{m}}^{2}\right|\boldsymbol{x}\right]}\right.\right.\\
 & \left.\left.\left.\vphantom{\left(\frac{1}{m}\left\Vert \hat{\boldsymbol{\mu}}_{k}-\boldsymbol{y}_{k}\right\Vert _{\mathbb{R}^{m}}^{2}\right.}-\mathbb{E}_{Y}\left[\left.\frac{1}{m}\left\Vert \hat{\boldsymbol{\mu}}_{k}-\boldsymbol{y}_{k}\right\Vert _{\mathbb{R}^{m}}^{2}-\frac{1}{m}\left\Vert \boldsymbol{\varepsilon}\right\Vert _{\mathbb{R}^{m}}^{2}\right|\boldsymbol{x}\right]\right)^{2}\right|\boldsymbol{x}\right]
\end{align*}
has been previously unknown, and 
the empirical variance of the ($K$-fold CV) loss is known to severely underestimate the variability induced by training-set randomness when samples overlap.
This makes the explicit expression of the
variance, and thus the tradeoff, unknown.

We derive its exact bound when it is available (symmetric error) and its upper bound otherwise (asymmetric error).
Specifically, our interest is in bounding the variance of the unbiased estimator of the pure loss, since we care about model correctness rather than data randomness.
The upper bound for the latter case serves as the upper limit on what the variance of the predictive accuracy can be, and is considered the worst-case scenario.

We first provide the exact bound for the case in which the errors, $\varepsilon_{i}$, are symmetric, though not necessarily identically distributed, letting $\sigma_{j}^{2}=\mathbb{E}\left[\varepsilon_{j}^{2}\right]$.
\begin{thm}\label{thm:var} Assume that $\varepsilon_{i}$
is symmetrically distributed. Then, the variance of the loss is,
\begin{alignat}{1}\label{eq:bound}
 & \mathbb{E}_{Y}\left[\left(\frac{1}{m}\left\Vert \hat{\boldsymbol{\mu}}_{k}-\boldsymbol{y}_{k}\right\Vert _{\mathbb{R}^{m}}^{2}-\frac{1}{m}\left\Vert \boldsymbol{\varepsilon}\right\Vert _{\mathbb{R}^{m}}^{2}\vphantom{\mathbb{E}_{Y}\left[\left.\frac{1}{m}\left\Vert \hat{\boldsymbol{\mu}}_{k}-\boldsymbol{y}_{k}\right\Vert _{\mathbb{R}^{m}}^{2}-\frac{1}{m}\left\Vert \boldsymbol{\varepsilon}\right\Vert _{\mathbb{R}^{m}}^{2}\right|\boldsymbol{x}\right]}\right.\right.\nonumber \\
 & \left.\left.\left.\vphantom{\left(\frac{1}{m}\left\Vert \hat{\boldsymbol{\mu}}_{k}-\boldsymbol{y}_{k}\right\Vert _{\mathbb{R}^{m}}^{2}\right.}-\mathbb{E}_{Y}\left[\left.\frac{1}{m}\left\Vert \hat{\boldsymbol{\mu}}_{k}-\boldsymbol{y}_{k}\right\Vert _{\mathbb{R}^{m}}^{2}-\frac{1}{m}\left\Vert \boldsymbol{\varepsilon}\right\Vert _{\mathbb{R}^{m}}^{2}\right|\boldsymbol{x}\right]\right)^{2}\right|\boldsymbol{x}\right]\nonumber \\
= & 4\frac{1}{m^{2}}\sum_{j\in C_{k}}\sigma_{j}^{2}\left(\hat{f}_{-k}(x_{j})-f(x_{j})\right)^{2}
\end{alignat}
if 
\[
\begin{cases}
=4\frac{1}{m^{2}}\sigma^{2}\sum_{j\in C_{k}}\left(\hat{f}_{-k}(x_{j})-f(x_{j})\right)^{2} & \left(\textrm{Homoskedastic}\right)\\
<4\frac{1}{m^{2}}\sigma_{\max_{j}}^{2}\sum_{j\in C_{k}}\left(\hat{f}_{-k}(x_{j})-f(x_{j})\right)^{2} & \left(\textrm{Heteroskedastic}\right)
\end{cases}
\]
\end{thm}
The right-hand side of \eqref{eq:bound} is what bounds the variance.
In the case of homoskedastic error, the bound is exact, meaning that we have the exact variance. 
If the error is heteroskedastic, the bound is an upper bound due to the variation in $\sigma^2_j$.
Intuitively, Theorem~\ref{thm:var} states that as the test-set size increases, the uncertainty in model evaluation shrinks.
When applied to $K$-fold cross-validation, i.e., an average of correlated hold-out losses, the bound remains valid but is necessarily conservative, as averaging folds reduces variance.
The proof is in Appendix~A.5.

Note that the bound depends on three variables: the model bias, $\frac{1}{m}\left(\sum_{j\in C_{k}}\left(\hat{f}_{-k}(x_{j})-f(x_{j})\right)^{2}\right)$, irreducible error, $\sigma^{2}$, and the hold-out size, $m$.
Intuitively, this theorem tells us that when test samples are more numerous (larger $m$), the uncertainty in our estimate of predictive accuracy shrinks due to the extra $\frac{1}{m}$, though at the cost of model accuracy from fewer training samples.

The model bias is the difference between the prediction and the data minus irreducible error, and can be estimated from the empirical cross-validation loss, given the specification of the irreducible error.

In many real datasets, errors may not be symmetrically distributed (e.g., binary outcomes), or evenly distributed (e.g., income data vary more among high earners). Theorem~\ref{thm:var2} relaxes the earlier assumptions to cover such cases.

When $\varepsilon$ is not symmetric, 
we have the following upper bound:
\begin{thm}\label{thm:var2}
Let $\sigma_{j}^{2}=\mathbb{E}\left[\varepsilon_{j}^{2}\right]$, we have
\begin{alignat*}{1}
 & \mathbb{E}_{Y}\left[\left(\frac{1}{m}\left\Vert \hat{\boldsymbol{\mu}}_{k}-\boldsymbol{y}_{k}\right\Vert _{\mathbb{R}^{m}}^{2}-\frac{1}{m}\left\Vert \boldsymbol{\varepsilon}\right\Vert _{\mathbb{R}^{m}}^{2}\vphantom{\mathbb{E}_{Y}\left[\left.\frac{1}{m}\left\Vert \hat{\boldsymbol{\mu}}_{k}-\boldsymbol{y}_{k}\right\Vert _{\mathbb{R}^{m}}^{2}-\frac{1}{m}\left\Vert \boldsymbol{\varepsilon}\right\Vert _{\mathbb{R}^{m}}^{2}\right|\boldsymbol{x}\right]}\right.\right.\\
 & \left.\left.\left.\vphantom{\left(\frac{1}{m}\left\Vert \hat{\boldsymbol{\mu}}_{k}-\boldsymbol{y}_{k}\right\Vert _{\mathbb{R}^{m}}^{2}\right.}-\mathbb{E}_{Y}\left[\left.\frac{1}{m}\left\Vert \hat{\boldsymbol{\mu}}_{k}-\boldsymbol{y}_{k}\right\Vert _{\mathbb{R}^{m}}^{2}-\frac{1}{m}\left\Vert \boldsymbol{\varepsilon}\right\Vert _{\mathbb{R}^{m}}^{2}\right|\boldsymbol{x}\right]\right)^{2}\right|\boldsymbol{x}\right]\\
\leqq & 16\frac{1}{m^{2}}\left(\sum_{j\in C_{k}}\sigma_{j}^{2}\left(\hat{f}_{-k}(x_{j})-f(x_{j})\right)^{2}\right)
\end{alignat*}
\end{thm}
If the error is not symmetric, the bound worsens as the constant term goes from $4$ to $16$.
Apart from that change,  the intuition is identical to Theorem~\ref{thm:var}.
The proof of this theorem is given in Appendix~A.6.

Together, the two theorems cover a broad set of problems, while the former provides an improved bound for a more specific case. 

One point to note is how correlations amongst folds affect this bound.
Although different folds in $K$-fold cross-validation are correlated because they share much of the same training data, this correlation cannot make the variability of the averaged $K$-fold estimate larger than the variability of any single fold. 
Formally, for fold losses $\{\hat L_1,\dots,\hat L_K\}$,
$\mathrm{Var}\!\left(\frac{1}{K}\sum_{k=1}^K \hat L_k\right)\le \max_k \mathrm{Var}(\hat L_k)$ (see Appendix~A.7).
Averaging several estimates can only keep the variance the same or reduce it, regardless of how strongly the folds overlap. 
Therefore, the bound we derive for the variability of a single evaluation fold is automatically a conservative bound for the full $K$-fold cross-validation estimate as well. In other words, even in the presence of dependence across folds, the upper bound we provide remains valid and safely overestimates the true uncertainty of the $K$-fold procedure.

Simulation results, including the complete hold-out variance analysis (Theorem~\ref{thm:var} and \ref{thm:var2}), $K$-fold variance comparisons, and all data-generating details, are presented in Appendix~A.2--4.

\section*{Determining the optimal hold-out size}

With the loss and its variance bound, we have an explicit tradeoff. To determine the optimal point of tradeoff,
we adopt the mean-variance utility within the von Neumann-Morgenstern
expected utility framework: 
\begin{align}
\textrm{Utility}\left(n,m\right)=-\left(\underbrace{\mathbb{E}\left[\hat{L}_{k}\left(\hat{\mu}_{k}\right)\right]}_{\textrm{Loss}}+\underbrace{\mathbb{V}\textrm{ar}\left[\hat{L}_{k}\left(\hat{\mu}_{k}\right)\right]}_{\textrm{Risk}}\right).\label{utility}
\end{align}
Here, this `utility' simply measures the overall quality of a split. A higher utility means a better balance between accuracy and uncertainty.
While there are other forms of utility, we select this utility (with unit weight on the variance term) due
to its simplicity and ease of interpretation, particularly how it is
analogous to the bias-variance tradeoff.
Because our variance bound scales linearly in $\sigma^2$ (and as $1/m$), introducing an additional penalty weight $\lambda$, as is done in typical mean-variance utility, would simply rescale $\sigma^2$ (i.e., it is equivalent to replacing $\sigma^2$ by $\lambda\sigma^2$).
Alternative risk measures could be substituted without affecting the role of the variance bound in determining the optimal hold-out size.

Based on the above, our proposed strategy for determining the optimal hold-out size involves four main steps:
\begin{itemize}
    \item[\textbf{Step 1:}] Estimate the model’s empirical prediction loss for several training/test splits;
    \item[\textbf{Step 2:}] Estimate how uncertain the losses in Step 1 are from the bound in Theorem~\ref{thm:var} or \ref{thm:var2};
    \item[\textbf{Step 3:}] Combine both into a single `utility' measure;
    \item[\textbf{Step 4:}] Choose the hold-out size that maximizes utility (the best tradeoff), which implies $K$.
\end{itemize}

To set up the above strategy, we first prepare the expected negative utility under mean-variance:
\begin{alignat*}{1}
 & \mathbb{E}_{X,Y}\left[\frac{1}{m}\left\Vert \hat{\boldsymbol{\mu}}_{k}-\boldsymbol{y}_{k}\right\Vert _{\mathbb{R}^{m}}^{2}-\frac{1}{m}\left\Vert \boldsymbol{\varepsilon}\right\Vert _{\mathbb{R}^{m}}^{2}\right]\\
 & \phantom{\mathbb{E}_{X,Y}}+\mathbb{E}_{Y}\left[\left(\frac{1}{m}\left\Vert \hat{\boldsymbol{\mu}}_{k}-\boldsymbol{y}_{k}\right\Vert _{\mathbb{R}^{m}}^{2}-\frac{1}{m}\left\Vert \boldsymbol{\varepsilon}\right\Vert _{\mathbb{R}^{m}}^{2}\vphantom{\mathbb{E}_{Y}\left[\left.\frac{1}{m}\left\Vert \hat{\boldsymbol{\mu}}_{k}-\boldsymbol{y}_{k}\right\Vert _{\mathbb{R}^{m}}^{2}-\frac{1}{m}\left\Vert \boldsymbol{\varepsilon}\right\Vert _{\mathbb{R}^{m}}^{2}\right|\boldsymbol{x}\right]}\right.\right.\\
 & \left.\left.\left.\phantom{\left(\left\Vert \hat{\boldsymbol{\mu}}_{k}\right\Vert _{\mathbb{R}^{m}}^{2}\right.}-\mathbb{E}_{Y}\left[\left.\frac{1}{m}\left\Vert \hat{\boldsymbol{\mu}}_{k}-\boldsymbol{y}_{k}\right\Vert _{\mathbb{R}^{m}}^{2}-\frac{1}{m}\left\Vert \boldsymbol{\varepsilon}\right\Vert _{\mathbb{R}^{m}}^{2}\right|\boldsymbol{x}\right]\right)^{2}\right|\boldsymbol{x}\right],
\end{alignat*}
where the first term is the loss and the second term is the uncertainty.
To calculate the above, we first need to estimate the two terms as
a function of $m$.

\textbf{(Step 1)} The first term (prediction loss) can be estimated empirically using
the empirical cross-validation loss: 
\begin{align}
\mathbb{E}\left[\left.\hat{L}_{k}\left(\hat{\mu}_{k}\right)\right|\boldsymbol{x}\right]\sim\frac{1}{m}\left\Vert \hat{\boldsymbol{\mu}}_{k}-\boldsymbol{y}_{k}\right\Vert _{\mathbb{R}^{m}}^{2}-\sigma^{2},
\end{align}
where $\sigma^{2}$ is the irreducible error. 
To determine this loss
as a function of $m$, or find the ``loss curve,'' we need the empirical loss at (at least) three anchor points: a very small test size ($m\approx 0$), a very large test size ($m\approx N$), and one intermediate value ($0<m<N$).
These reference points anchor the curve that relates test-set size to model performance, allowing us to interpolate how loss changes with $m$.
Because $m=0$ is not operational (no evaluation set; though can be mathematically derived, discussed later), we approximate this endpoint using leave-one-out CV (LOOCV; $m=1$), which provides an empirical proxy for the ``minimal hold-out'' regime.
Estimating the loss when $m=N$ is also possible, but it
is unrealistic that the optimal test size lies near $n=0$, so instead
we use some minimal number for the model to produce reasonable predictions
(in our example, we use $K=2$), which we call leave-most-out cross-validation (LMOCV). Finally, we calculate some $0<m<N$, e.g., 5-fold
CV. Each loss is denoted as $\hat{L}_{LOO}$, $\hat{L}_{LMO}$, and
$\hat{L}_{5-fold}$, and sample size as $m_{LOO}$, $m_{LMO}$, and
$m_{5-fold}$, respectively. 
Since the loss is monotone increasing,
the curve can be estimated as $E_{\hat{L}}(m)=m^{\frac{\log(\beta)}{\log(\alpha)}}({\hat{L}_{LMO}-\hat{L}_{LOO}})+\hat{L}_{LOO}-\sigma^{2}$,
where $\beta=\frac{\hat{L}_{5-fold}-\hat{L}_{LOO}}{\hat{L}_{LMO}-\hat{L}_{LOO}}$
and $\alpha=\frac{m_{5-fold}-m_{LOO}}{m_{LMO}-m_{LOO}}$. While there
are other ways to estimate this curve, the above empirically fits
well and satisfies all constraints (a simulation study that confirms this is given in Appendix~A.1, which also shows that the curve is robust against the choice of smoother). From this, we have the loss curve.
Note that the estimation of the loss curve can be avoided entirely if the researcher is only interested in finding the optimal $K$ within a limited set of $K$ (e.g., $K=4,5,10,20$).

\textbf{(Step 2)} Given the loss curve, $E_{\hat{L}}(m)$, the bound of the variance
curve, $V_{\hat{L}}(m)$, is calculated as: 
\begin{align*}
V_{\hat{L}}(m)=4\frac{1}{m}\sigma^{2}E_{\hat{L}}(m)
\end{align*}
Again, $\sigma^{2}$ is the irreducible noise of the data. 
This formula shows how uncertainty decreases as test-set size increases, providing a way to quantify the tradeoff illustrated earlier.

\textbf{(Step 3)} Finally, the utility can be calculated as, 
\begin{alignat*}{1}
\textrm{Utility(m)}= & -\left(E_{\hat{L}}(m)+V_{\hat{L}}(m)\right).
\end{alignat*}
\textbf{(Step 4)} Given $\sigma^{2}$, the optimal test set size is what maximizes this
utility (or minimizes $E_{\hat{L}}(m)+V_{\hat{L}}(m)$).

\section*{Making implicit assumptions explicit: the role of $\sigma^{2}$}

A central element of our approach-- and of any cross-validation procedure-- is the irreducible noise of the data-generating process, $\sigma^{2}$. 
Every choice of hold-out size implicitly assumes some value of $\sigma^{2}$, and conventional approaches leave this assumption unexamined. 
For example, choosing 80/20 for a particular dataset and model implies a specific belief about this noise that differs from 90/10. 
Our framework makes this implicit assumption explicit and operational.

In practical terms, $\sigma^{2}$ reflects how noisy or unpredictable the data are. In physical sciences, $\sigma^{2}$ may be small (e.g., controlled experiments), whereas it can be large in social sciences.
For example, in a typical biological experiment, where measurement variability is roughly 10\% of total signal, a good estimate could be $\sigma^{2}\approx0.1$, while in a social survey, responses vary widely; $\sigma^{2}\approx1-2$.
Precisely identifying $\sigma^2$ would require strong knowledge about the data-generating process.
In practice, $\sigma^2$ is rarely known and is better treated as an assumed noise level (or explored
via sensitivity analysis), which is exactly what our procedure makes explicit.
While the ability to choose $\sigma^{2}$ may seem too subjective, it is important to note that the researcher is implicitly assuming some $\sigma^{2}$ whenever they choose a hold-out size.
Rather than viewing this as a limitation, this framing allows for the researcher to conduct a useful sensitivity analysis, or ``reverse engineer'' implicit noise levels under conventional hold-out splits.
In other words, $\sigma^{2}$ is a ``knob'' that the researcher can ``tune,'' allowing for the researcher to take an implicit assumption and relax it by exploring other assumptions.
For example, by varying $\sigma^{2}$ over a plausible range, researchers can examine how the optimal test set changes, and identify robustness ranges where splits remain near-optimal (e.g., “for all 
$\sigma^{2}\in[0.5,2.0]$, 5-fold CV remains near-optimal”).
Reverse engineering $\sigma^{2}$ allows the researcher to interpret the choice of 
implied $K$ in terms of underlying assumptions about data quality.
Both uses provide ways to understand $\sigma^{2}$ from a predictive perspective.
Since the researcher is already assuming some $\sigma^{2}$ when they conduct $K$-fold CV, treating $\sigma^{2}$ as a tuning parameter allows the researcher to reveal their implicit assumptions and explore other assumptions.

A more holistic view is to examine how the optimal test size changes given a wide range of $\sigma^{2}$.
This relation defines the Pareto frontier, which summarizes all optimal tradeoffs between accuracy and uncertainty for different assumed noise levels.
The Pareto frontier is analogous to modern portfolio theory, where one specifies the risk aversion parameter, $\lambda$, to calculate the efficient frontier of the optimal portfolio (i.e., what is the optimal portfolio proportion given risk aversion).
Thus, choosing $\sigma^{2}$ is similar to choosing between high-risk, high-return and low-risk, low-return investments.
If the researcher wishes to prioritize reducing uncertainty in the estimated prediction loss,
they should examine larger (more conservative) values of $\sigma^2$; if they are less concerned
about uncertainty, smaller values are appropriate.
Note that, when calculating the Pareto frontier, there is a point where $\sigma^{2}$ is too large and the frontier cuts off.
This is an-- albeit loose-- upper bound on $\sigma^{2}$, calculated from the model and data.
This provides some idea regarding what the most risk-averse $\sigma^{2}$ is, from which we can calculate the ``safest'' split for cross-validation.

If necessary, using the MSE of a low-bias estimator as a data-driven anchor is possible, though can be misleading if the discrepancy is large between the true $\sigma^2$.

For heteroskedastic errors, one conservative choice is to use an upper bound on the noise level, e.g.\ $\sigma^2=\max_j \sigma_j^2$, which preserves the bound at the cost of additional conservatism.

\section*{Real-world example: Abalone age prediction}

To demonstrate our approach in practice, we apply it to a regression problem using the Abalone dataset \citep{abalone_1}.
The Abalone dataset has $N=4,177$ observations with eight covariates, and the goal is to predict the age of the abalone based on these features.
We consider two models, a linear regression model and a random forest model, for comparison to estimate the optimal test set size, due to the latter being a more complex (nonlinear) model compared to the former.

\subsection*{Step 1: Estimate loss curve}
First, we calculate the LOOCV squared error loss, $\hat{L}_{LOO}$, by removing one sample for testing ($m=1$) and using the rest ($n=4,176$) to train the models.
This produces a loss value of 4.9394 for the linear model and 4.6379 for the random forest.

Second, we calculate the LMOCV squared error loss, $\hat{L}_{LMO}$, by removing half of the samples ($m=2,088$) for testing and using the rest ($n=2,089$) to train the models.
This produces a loss value of 4.9594 for the linear model and 5.0571 for the random forest.

Third, we calculate the 5-fold CV squared error loss, $\hat{L}_{5-fold}$, by removing 1/5th of samples ($m=835$) for testing and using the rest ($n=3,342$) to train the models.
This produces a loss value of 4.9426 for the linear model and 4.6692 for the random forest.

Plugging this into the loss curve function, we obtain $E_{\hat{L}}(m)=m^{2.0010}0.0200+4.9394-\sigma^2$ for the linear model and $E_{\hat{L}}(m)=m^{2.7898}0.4192+4.6379-\sigma^2$ for the random forest.

\begin{figure}[t!]
\centering
\includegraphics[width=0.5\textwidth,clip]{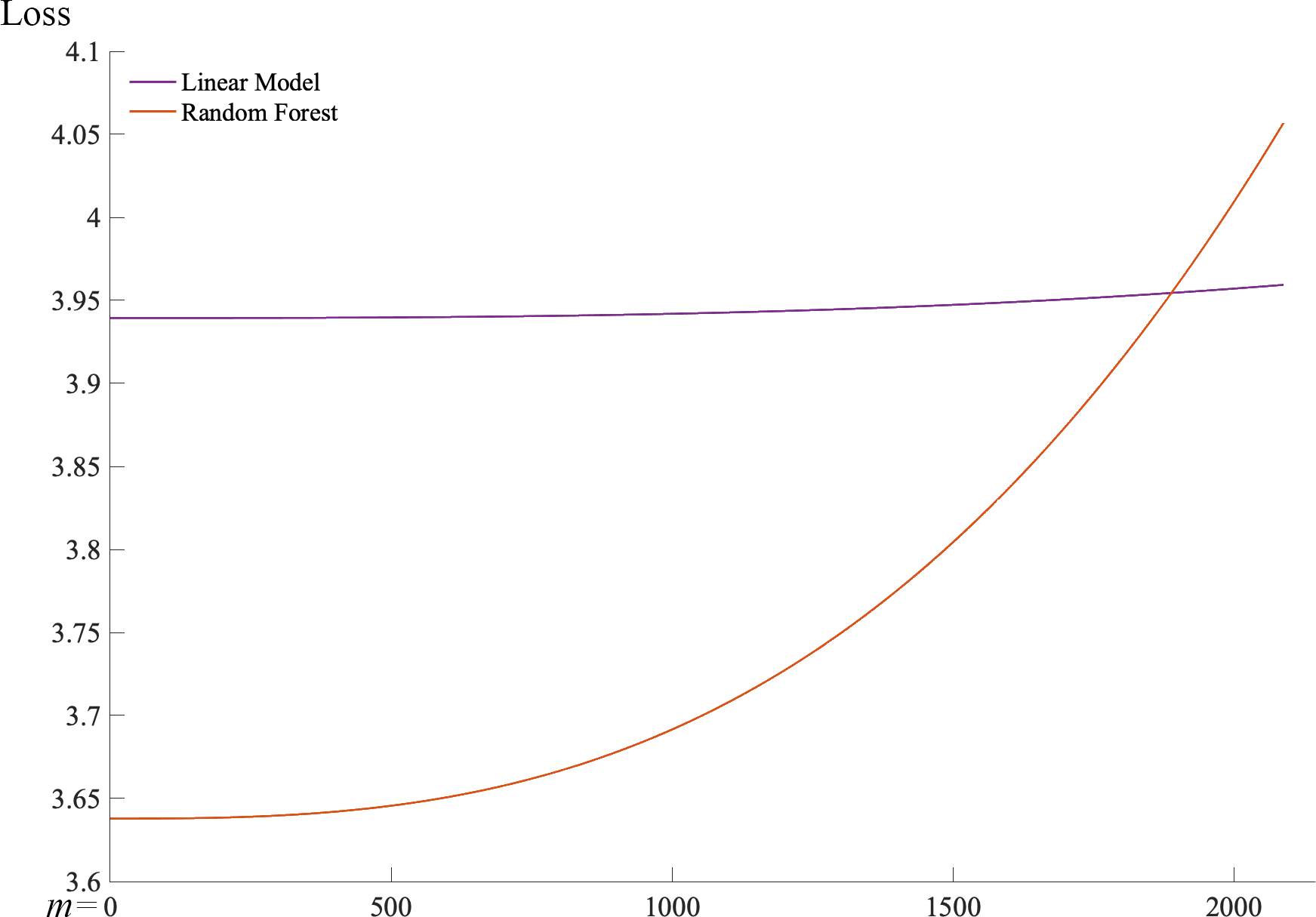}
\caption{Abalone data: Estimated loss curve for the linear model and random forest. The y-axis is the estimated mean squared error under different hold-out data sizes ($m$: x-axis). $\sigma^2$ is the irreducible error, which is set to $\sigma^2=1$ for illustration. 
\label{fig:loss}
}
\end{figure}

The estimated loss curve, with $\sigma^2=1$, is visualized in Figure~\ref{fig:loss}, which shows how prediction loss changes with the size of the test set (we set $\sigma^2=1$ purely as a normalization for this illustrative example; the recommended hold-out size is evaluated over a range of $\sigma^2$ values via the Pareto frontier and robustness ranges below).
Because increasing $m$ reduces the training set size, we expect the prediction loss to increase (typically smoothly) with $m$.
The flatness of the curve for the linear model means its performance stabilizes quickly, while the steeper curve for the random forest shows that complex models need more training data.

\subsection*{Step 2: Calculate variance curve}

After assessing how model accuracy changes, we next examine how the uncertainty of this evaluation behaves.
Given the estimated loss curve in Step 1, we can directly calculate the bound of the variance curve as
\begin{align*}
    V_{\hat{L}}(m)=4\frac{1}{m}\sigma^2E_{\hat{L}}(m).
\end{align*}

\begin{figure}[t!]
\centering
\includegraphics[width=0.5\textwidth,clip]{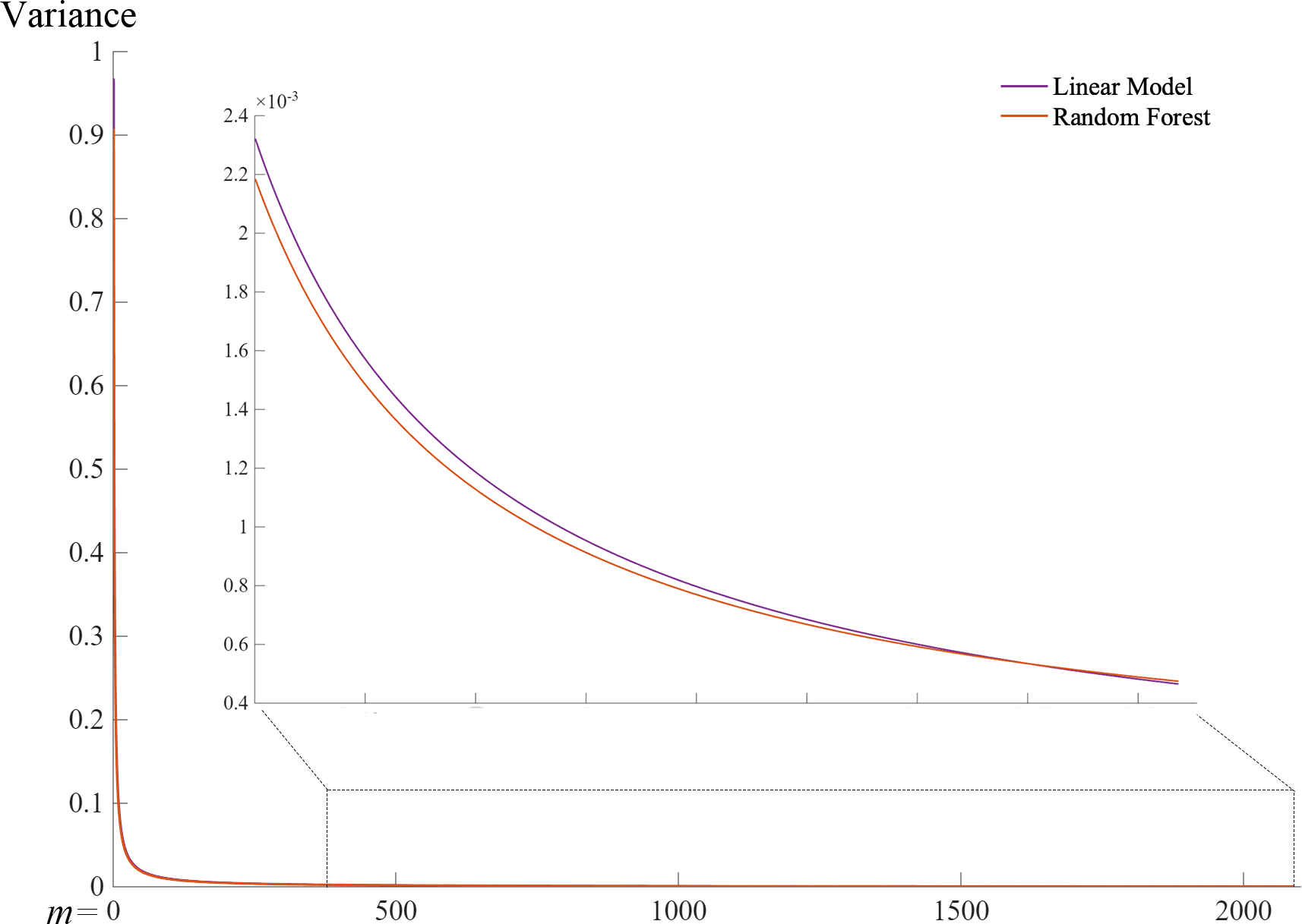}
\caption{Abalone data: Estimated variance curve for the linear model and random forest. The y-axis is the calculated bound of the variance of the loss under different hold-out data sizes ($m$: x-axis). $\sigma^2$ is the irreducible error, which is set to $\sigma^2=1$ for illustration. 
\label{fig:var}
}
\end{figure}

The calculated variance curve, with $\sigma^2=1$, is visualized in Figure~\ref{fig:var}.
The variance is extremely high when $m=1$, because the evaluation is based on a single data point, and there is a large amount of uncertainty regarding whether this one sample is representative of the population.
As $m$ increases, the uncertainty decreases, as there are more data to assess the predictive ability of a model.
Compared to the loss curve, the variance curve is less contrastive, though it follows a similar pattern to the loss curve in terms of the difference between the two models.

\subsection*{Step 3: Calculate the mean-variance utility}
Given both the loss curve and the variance curve, we can calculate the negative utility,
\begin{align*}
E_{\hat{L}}(m) +V_{\hat{L}}(m),
\end{align*}
and whatever minimizes this value, given some value of  $\sigma^2$, is the optimal test set size.

\begin{figure*}[t!]
\centering
\includegraphics[width=1\textwidth,clip]{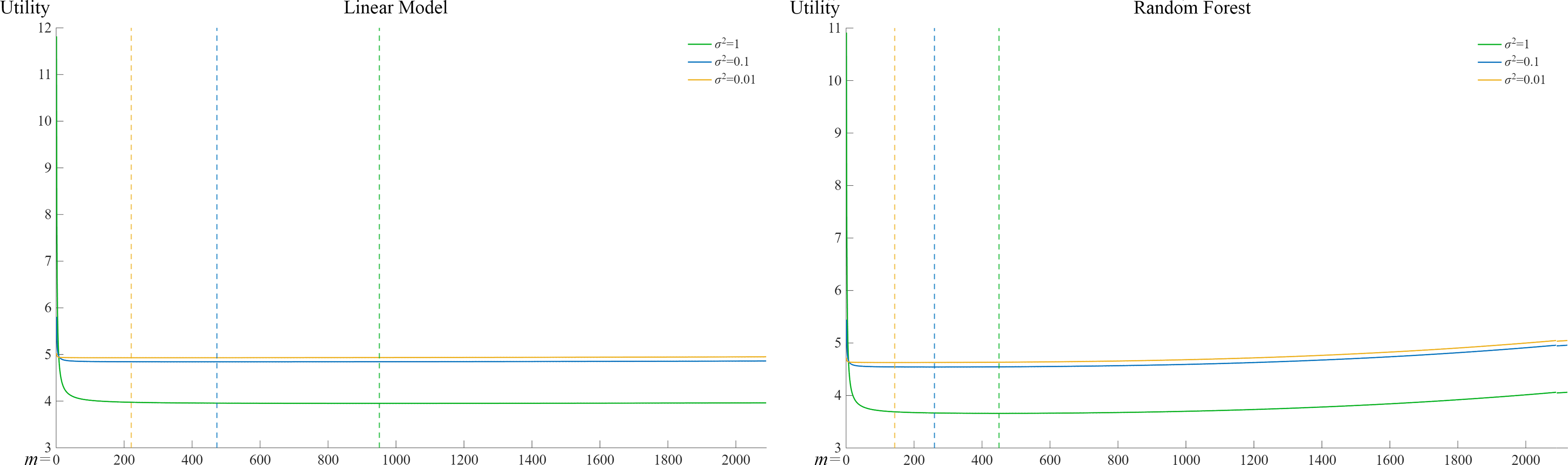}
\caption{Abalone data: Negative utility of the linear model (left) and random forest (right), for $\sigma^2=\{0.01,0.1,1\}$. Dashed lines are the minimal point under each $\sigma^2$.
\label{fig:lambda}
}
\end{figure*}

Figure~\ref{fig:lambda} summarizes how the overall utility changes with test-set size under different noise assumptions ($\sigma^2=\{0.01,0.1,1\}$). The lowest point on each curve marks the best choice of $K$, as it minimizes the negative utility (and maximizes the utility).
For the linear model, the optimal test set ranges between roughly 200 and 950 samples; for the random forest, between 150 and 450.
The optimal test set size is different between the two models, reflecting the difference in model complexity.
In particular, the linear model is a faster learner, requiring fewer data for training and allocating more for testing.
The optimal test size of the linear model also varies more than the random forest given the specification of $\sigma^2$.
Notably, except for the linear model with $\sigma^2=1$, neither model has the optimal test size around $K=5$.
Given these results, a researcher using the random forest with this data should not be using $K=5$ because the model has not learned enough for accurate predictions.

\subsection*{Step 4: Calculate the Pareto frontier}

\begin{figure}[t!]
\centering
\includegraphics[width=0.5\textwidth,clip]{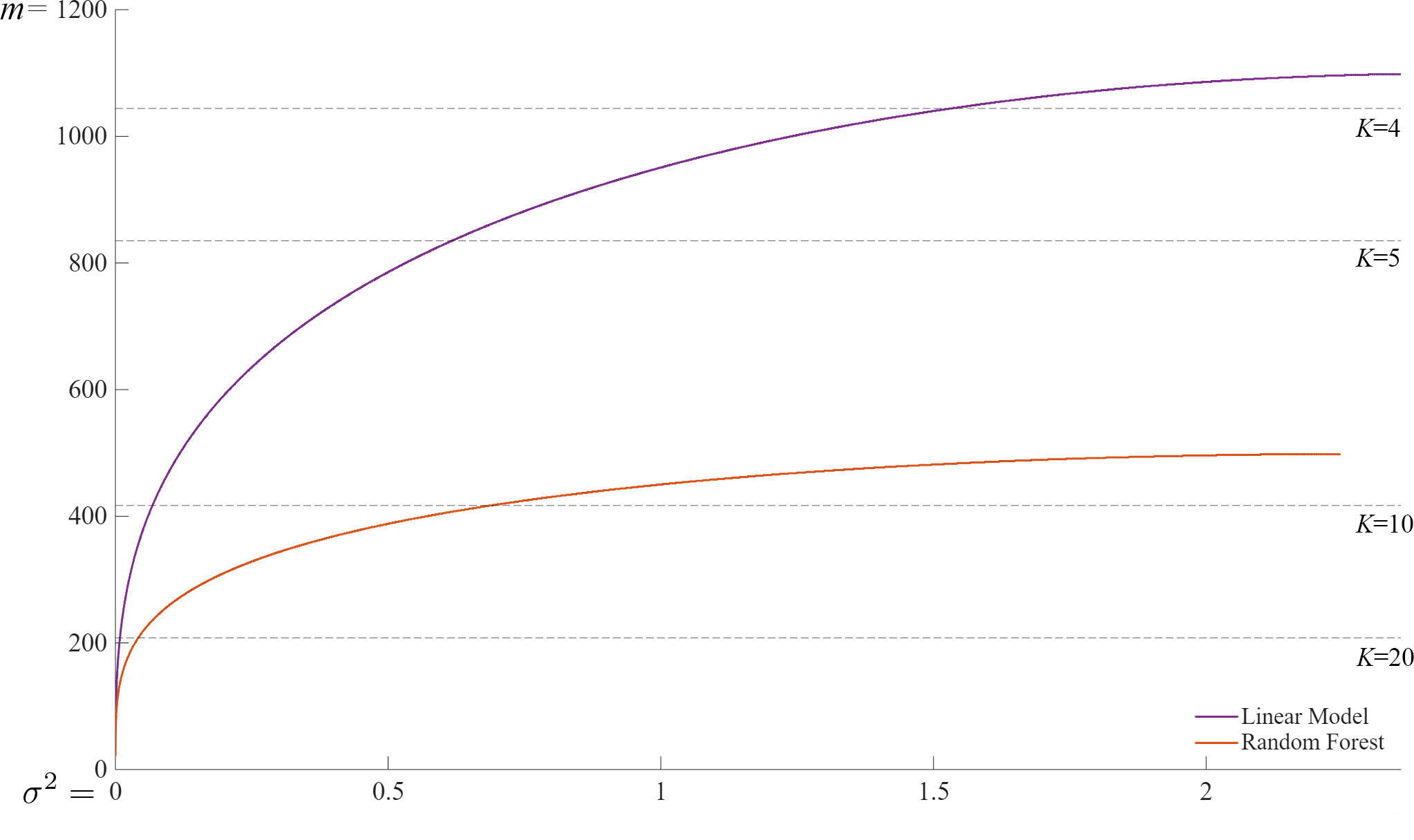}
\caption{Abalone data: Pareto frontier for the linear model and random forest. Dashed lines denote sample size that equals $K=4,5,10,20$, with $m=N/K$.
\label{fig:front}
}
\end{figure}

Once the utility is calculated, we can obtain the Pareto frontier by varying $\sigma^2$.
This is done by calculating the minimal test set size under $\sigma^2$, while varying $\sigma^2$  until the minimal test set size peaks.
The Pareto frontier for this data set is given in Figure~\ref{fig:front}.

The frontier maxes out at around $\sigma^2=2.25\sim 2.5$, which is the loose upper bound of the irreducible noise.
Comparing the two models, we see that the random forest has a much gradual curve than the linear model.
This shows that, no matter what the belief of $\sigma^2$ is, the random forest requires a certain (large) amount of data for training.
The linear model, with its steep curve, shows that, because the model is simple enough to require fewer data for training, the optimal sample size is sensitive to the specification of $\sigma^2$.
As a result, $K=4,5$ is never optimal for the random forest, but is for the linear model if we assume that $\sigma^2=0.7\sim 1.6$.
The linear model is a faster learner and requires fewer data to be ``certain'' compared to the more complex random forest.
We can then calculate the implicit $\sigma^2$ under different $K$, or analyze how different models learn the data.

The key takeaway is that, even for a common dataset, the optimal hold-out size, and thus implied $K$, varies between models.
Using a conventional split for both models, thus, can mislead model evaluation if the implicit assumption of $\sigma^2$ is drastically different.

\section*{Comparison with different datasets}
To further illustrate how the optimal test size differs by model and dataset, we expand the analysis to five benchmark datasets  (including the Abalone data in the previous section) that span multiple domains.
The descriptions of datasets are given in Table~\ref{table: dataset}.
The datasets span diverse domains, from engineering, biology, to social science, with total sample sizes from 167 to over 32,561. This range enables us to compare how optimal test-set sizes differ when data noise levels vary across fields.

\begin{table*}[t!]
\tiny
\centering
    \caption{Description of datasets used to compare optimal hold-out size. $N$ is the total sample size and $p$ is the number of covariates}
    \label{table: dataset}
\begin{tabular}{lllllrr}
\hline
 & Description & Source & Domain & Target & N & p \\ \hline
\texttt{Abalone}: & Age of abalone &\cite{abalone_1}  & Biology  & Rings & 4,177 & 8 \\
\texttt{Concrete}: & Concrete compressive strength &\cite{concrete_compressive_strength_165}  & Materials Engineering & Concrete strength & 1,030 & 8 \\
\texttt{Servo}: & Simulation of a servo system &\cite{servo_87}  & Engineering & Rise time & 167 & 4 \\
\texttt{Diabetes}: & Diabetes study & \cite{efron2004least} & Medicine & Disease progression score & 442 & 10 \\
\texttt{Income}: & Census income & \cite{adult_2} & Social Science & log(Income) & 32,561 & 5\\ \hline
\end{tabular}
\end{table*}

Table~\ref{table: optimal} summarizes the optimal test-set sizes across five datasets, two models, and three noise levels.
Overall, the observations in the Abalone application are consistent for the other datasets: the optimal test set size increases as $\sigma^2$ increases, and the linear model requires fewer training data than the random forest (because the linear model is a faster learner).
The exception to this observation is with the \texttt{Diabetes} dataset, where the random forest requires more test data than the linear model (for $\sigma^2=0.01$).

Note that there were instances where the empirical loss was less than $\sigma^2=1$, denoted by ``-'' (for \texttt{Diabetes}, RF, and \texttt{Income}, LM and RF).
These entries provide valuable diagnostic information that the data are telling us that $\sigma^2=1$ is inconsistent with the observed model performance. 
If a researcher truly believed $\sigma^2\geq1$ for these datasets, they should question whether their models are overfitting or whether their noise assumptions are incorrect. 
For valid analysis, one should only consider $\sigma^2<1$. 
This diagnostic capability is precisely what conventional $K$ selection obscures.

An important result is that, for all datasets and models considered, the optimal test-set size is neither leave-one-out (LOOCV), nor 2-fold.
In most cases, the optimal hold-out size, in terms of $K$, is in the range of $4-20$, depending on the model and assumption on $\sigma^2$.
This is somewhat in line with conventional wisdom, though the variability given the data, model, and $\sigma^2$ is quite stark.
The results show that the best hold-out size varies widely depending on the dataset and model, ranging roughly from 4 to 20 folds. This variability highlights why a fixed choice (e.g., 5-fold CV) can be misleading.
In practical terms, this means researchers should test several hold-out sizes rather than assume one default, especially when the dataset size or expected noise level differs across studies.

\begin{table}[t!]
\centering
    \caption{The optimal hold-out size for five datasets, for both linear model (LM) and random forest (RF), under varying noise levels, $\sigma^2=\{0.01,0.1,1\}$. Entries with ``-'' indicate that the estimated loss was smaller than the irreducible noise.}
    \label{table: optimal}
\begin{tabular}{lrrrrr}
\hline
 & N & \multicolumn{1}{r}{Model} & $\sigma^2$=   0.01 & 0.1 & 1 \\ \hline
\multirow{2}{*}{\texttt{Abalone}} & \multirow{2}{*}{4,177} & LM & 221 & 473 & 951 \\
 &  & RF & 143 & 260 & 450 \\
\multirow{2}{*}{\texttt{Concrete}} & \multirow{2}{*}{1,030} & LM & 44 & 114 & 299 \\
 &  & RF & 17 & 34 & 69 \\
\multirow{2}{*}{\texttt{Servo}} & \multirow{2}{*}{167} & LM & 6 & 16 & 21 \\
 &  & RF & 5 & 12 & - \\
\multirow{2}{*}{\texttt{Diabetes}} & \multirow{2}{*}{442} & LM & 16 & 46 & 133 \\
 &  & RF & 21 & 38 & 71 \\
\multirow{2}{*}{\texttt{Income}} & \multirow{2}{*}{32,561} & LM & 1,675 & 4,165 & - \\
 &  & RF & 1,412 & 2,053 & - \\ \hline
\end{tabular}
\end{table}

Finally, we reverse engineer to estimate the implicit $\sigma^2$, given a specific split, implying a specific $K$.
The results, with $K=\{4,5,10,20\}$, is reported in \ref{table: sigma}.
Overall, the results are in line with earlier results: the implied $\sigma^2$ varies amongst models and data, and for each $K$.
Within $K$, the implicit $\sigma^2$ for both models is somewhat consistent, particularly for large $K$.
However, this value differs drastically across different $K$.
Furthermore, for the random forest, the implicit $\sigma^2$ does not exist for $K=4,5$ (except for the \texttt{Diabetes} dataset), and does not exist for $K=10$ for the \texttt{Servo} and \texttt{Income} datasets.
This means that, if a researcher chooses those $K$ for those datasets, there does not exist an implied  $\sigma^2$, and thus the optimal hold-out size does not exist.
From a researcher's perspective, this narrows down the potential $K$ in a data-driven manner.
Reverse engineering $\sigma^2$, thus, greatly aids in determining $K$ by removing some $K$ from the candidates.
Given a smaller set of $K$, the researcher can select a $K$ that best fits their assumption regarding the noise.
These findings reinforce that optimal $K$ is context-specific and that reverse-engineering $\sigma^2$ can guide researchers toward data-consistent choices.

\begin{table}[t!]
\centering
    \caption{The implicit $\sigma^2$ for five datasets, for both linear model (LM) and random forest (RF), under varying hold-out size, implying $K=\{4,5,10,20\}$, with $m=N/K$. Entries with ``-'' indicate that the implied $\sigma^2$, under that $K$, is greater than the upper bound of $\sigma^2$, i.e., these $K$ values are too small for the model to learn effectively from these data.}
    \label{table: sigma}
\begin{tabular}{lrrrrr}
\hline
 & Model & $K=$ 4 & 5 & 10 & 20 \\ \hline
\multirow{2}{*}{\texttt{Abalone}} & LM & 1.5284 & 0.6160 & 0.0683 & 0.0084 \\
 & RF & - & - & 0.6847 & 0.0418 \\
\multirow{2}{*}{\texttt{Concrete}} & LM & 0.6960 & 0.4054 & 0.0770 & 0.0149 \\
 & RF & - & - & 4.1242 & 0.3783 \\
\multirow{2}{*}{\texttt{Servo}} & LM & - & - & 0.0908 & 0.0114 \\
 & RF & - & - & - & 0.0267 \\
\multirow{2}{*}{\texttt{Diabetes}} & LM & 0.6732 & 0.4061 & 0.0898 & 0.0196 \\
 & RF & 5.3339 & 2.1891 & 0.1602 & 0.0115 \\
\multirow{2}{*}{\texttt{Income}} & LM & - & - & 0.0493 & 0.0094 \\
 & RF & - & - & - & 0.0225 \\ \hline
\end{tabular}
\end{table}

\section*{Why $K$ matters scientifically}\label{sec:whyKmatters}

Mutational-signature attribution turns high-dimensional somatic mutation data into a sparse set of active
mutational processes, typically represented in the 96 trinucleotide-context basis \citep{Alexandrov2013}.
Because the scientific conclusions are expressed in terms of {which} signatures are deemed present and
{how} strongly they contribute, any tuning choice that changes sparsity can change downstream
mechanistic claims.

We illustrate this using TCGA/MC3 somatic mutation calls \citep{Ellrott2018MC3}, converted to SBS96 spectra
using standard MAF tooling \citep{Mayakonda2018maftools} and refit against the COSMIC reference catalogue
\citep{Sondka2024COSMIC}. Signature weights are estimated via $\ell_1$-regularized
(nonnegative) regression, as in signature ``refitting'' approaches designed to yield sparse, interpretable
assignments \citep{Li2020sigLASSO,Friedman2010glmnet}. We then apply our proposed utility framework:
for each fold count $K$, compute the hold-out loss and select the $K$ that minimizes the corresponding
risk-adjusted estimate of pure loss as a function of $\sigma^2$.

The results are reported in Table~\ref{tab:sigma2K}.
First, the {fit} can be nearly insensitive to $K$: across a wide range
of fold counts, the empirical hold-out loss changes only marginally. Second, despite that near-flat fit
curve, the {optimal} $K$ can vary sharply with $\sigma^2$ under the pure-loss criterion, producing a
step-function frontier $K$ in which larger uncertainty favors smaller $K$. This is exactly the
regime in which downstream selection is most fragile. When multiple $K$ values have essentially identical
predictive performance, the fold design becomes a lever on sparsity rather than accuracy.

In our TCGA analysis, this translates into different substantive conclusions about which signatures are
``reliably present.'' Using a stability threshold (a signature is declared present only if selected in at
least 70\% of tumors), the high-uncertainty regime (where the pure-loss criterion selects small $K$)
excludes signatures that clear the same threshold under lower uncertainty (where the criterion selects a
larger $K$). For example, SBS24, SBS39, and SBS49 cross the stability threshold when moving from the
high-uncertainty optimum ($K=2$) to the lower-uncertainty optima ($K\ge 5$). Thus, varying $K$ changes not
only numerical weights but the discrete set of inferred processes and, consequently, the biological
narratives supported by the data.

 \begin{table}[t]
 \centering
 \caption{Optimal $K$ as a function of $\sigma^2$ (TCGA MC3).}
 \label{tab:sigma2K}
 \small
 \begin{tabular}{rrr}
 \hline
 $K$ & $\sigma^2$ range (low) & $\sigma^2$ range (high) \\
 \hline
 100 & 0 & $7.41\times 10^{-4}$ \\
 20  & $7.41\times 10^{-4}$ & $6.84\times 10^{-3}$ \\
 10  & $6.84\times 10^{-3}$ & $4.03\times 10^{-2}$ \\
 5   & $4.03\times 10^{-2}$ & $3.52\times 10^{-1}$ \\
 2   & $3.52\times 10^{-1}$ & -- \\
 \hline
 \end{tabular}
\end{table}

\section*{Computational considerations}

While conceptually straightforward, the proposed method can be computationally demanding because it requires multiple rounds of model fitting.
At a minimum, cross-validation has to be performed three times, which can be too prohibitive for more complex machine learning methods, such as neural networks.
To decrease computation cost, one can avoid full cross-validation to approximate the prediction loss.
For example, a 5-fold CV can be assessed with only one test set, rather than the full five.
Although there are some accuracy concerns with this approximation, it is fairly robust as long as the test set is not wildly inconsistent with the training set.

Modern parallel computing alleviates many of these concerns, but some large-scale machine learning methods might still find this computationally prohibitive.
In some cases, one can use the theoretical convergence rate of the model as a stand-in.
The difficulty, however, is in deriving these rates for all of the models considered.
Another way is to use an approximate model that is less computationally intensive but retains the core characteristics.

The most computationally intensive cross-validation is the LOOCV.
As mentioned above, we can calculate the Stein's unbiased risk estimator (SURE) when $m=0$.
Since $\hat{\boldsymbol{\mu}}$ is generally a function of $\boldsymbol{y}$, $\hat{\boldsymbol{\mu}}\left(\boldsymbol{y}\right)$, from Stein's equality, we have
\[
\mathbb{E}\left[\left.\boldsymbol{\varepsilon}^{\top}\hat{\boldsymbol{\mu}}\left(\boldsymbol{x}\right)\right|\boldsymbol{x}\right]=\mathbb{E}\left[\left.\sigma^{2}\sum_{i=1}^{n+m}\frac{\partial\hat{\mu}}{\partial y_{i}}\left(\boldsymbol{x}_{i}\right)\right|\boldsymbol{X}\right].
\]
Then, the SURE is, 
\begin{align*}
\widehat{\textrm{SURE}}=\left\Vert \hat{\boldsymbol{\mu}}-\boldsymbol{y}\right\Vert _{\mathbb{R}^{n+m}}^{2}+2\sigma^{2}\sum_{i=1}^{n+m}\frac{\partial\hat{\mu}}{\partial y_{i}}\left(\boldsymbol{x}_{i}\right)-\sigma^{2}\left(n+m\right).
\end{align*}
The unbiased estimate of variance of the SURE (SURE for SURE) is
\begin{alignat*}{1}
\widehat{\textrm{SURE for SURE}} & =4\left\Vert \hat{\boldsymbol{\mu}}-\boldsymbol{y}\right\Vert _{\mathbb{R}^{n+m}}^{2}\\
 & \phantom{=4}+4\sigma^{2}\textrm{trace}\left\{ \left(\nabla\hat{\boldsymbol{\mu}}\left(\boldsymbol{y}\right)\right)^{2}\right\} -2\left(n+m\right)\sigma^{2},
\end{alignat*}
where
\[
\nabla\hat{\boldsymbol{\mu}}\left(\boldsymbol{y}\right)=\left[\begin{array}{ccc}
\frac{\partial\hat{\mu}_{1}}{\partial y_{1}}\left(\boldsymbol{y}\right) & \cdots & \frac{\partial\hat{\mu}_{n+m}}{\partial y_{1}}\left(\boldsymbol{y}\right)\\
\vdots & \ddots & \vdots\\
\frac{\partial\hat{\mu}_{1}}{\partial y_{n+m}}\left(\boldsymbol{y}\right) &  & \frac{\partial\hat{\mu}_{n+m}}{\partial y_{n+m}}\left(\boldsymbol{y}\right)
\end{array}\right].
\]
This formula provides an analytic shortcut to estimate prediction risk without retraining the model for every possible test split.
While $\textrm{div}$ can be calculated for simpler models (e.g., linear models), it is quite prohibitive for more complex models, such as neural networks.
Whether one uses LOOCV or SURE depends on the computational cost and mathematical availability.

\section*{Discussion}

The cross-validation procedure can be viewed as a balance between two competing goals: model accuracy and evaluation certainty.
Based on this tradeoff, we devise a strategy based on multi-objective optimization to determine the optimal hold-out size in cross-validation.
Our proposed strategy is easily implemented through basic cross-validation estimates and is optimized based on the researcher's belief in the irreducible noise.
Through regression exercises, we demonstrate how the optimal test set size can be estimated, and how that informs each model's predictive ability, sensitivity to assumptions regarding the irreducible error, and how conventional splits can mislead evaluation.
Our proposed method can be applied to many statistical problems, but is particularly useful (due to an exact identity) for regression problems where the errors are symmetrically distributed and homoskedastic.

For applied scientists, the main takeaway is conceptual: there is no universal best split size. Instead, it should be chosen based on data quality and model complexity, similar to how one adjusts sample size or significance levels in experimental design.

The fundamental contribution of this work is not eliminating the need for assumptions in cross-validation, which is impossible, but making existing hidden assumptions explicit and actionable.
Every choice of hold-out size in conventional practice implicitly embeds beliefs about the irreducible noise, model learning rates, and acceptable uncertainty levels.
Rather than relying on these implicit assumptions, our framework provides tools to align cross-validation design with domain knowledge and data characteristics.
This not only allows researchers to explicate and test their assumptions, but also improves transparency in scientific research.

Our approach has several limitations. The most important thing is the computational cost of estimating the cross-validation loss.
While we have discussed ways to reduce this cost, another approach is to use the convergence rate of each model, though this requires knowledge of this rate, which can be difficult to obtain.
As discussed above, whether one reduces the number of test sets, uses the convergence rate, or SURE to avoid LOOCV, depends on the availability of computational resources and theoretical understanding.

Finally, as seen in the results of our analysis, it is not always necessary to obtain exact optimal test set sizes.
Our results show that, for the Abalone data, there is a reasonable range for which little utility is lost by deviating from the optimal size.
The key result is that the optimal size will rarely be LOOCV, or something like a 2-fold CV, but somewhere in between.
Instead of relying on a single conventional choice across models and datasets, we recommend identifying a reasonable range of hold-out sizes tailored to specific problem types or data domains.
This is similar to having different significance levels for statistical testing for different domains.
The main takeaway is that there is no one-size-fits-all for cross-validation, and it should be considered on a domain-by-domain, dataset-by-dataset, and model-by-model basis.

As a general heuristic, our results indicate that the hold-out size should be small if the researcher believes that the irreducible noise is small (e.g., physics compared to economics), or the model is complex and requires more data for training.
Conversely, the hold-out size can be large, which has the benefit of reducing computation, if the irreducible noise is large or the model is simple.
The difficulty lies in the middle ground, e.g., when the noise is large but the model is complex, or when both the noise and model are moderate.
For these cases, the optimal hold-out size should be calculated, or, at the very least, a reasonable range of splits should be explored.

Our theoretical results provide an exact expression and upper bound for the evaluation uncertainty, depending on the error assumption, at a given hold-out size $m$, which is directly applicable to single-split validation and remains conservative when implemented through $K$-fold averaging. A natural direction for future work is to derive bounds that directly target the variance of the {$K$-fold averaged} estimator. Such results would require explicit control of cross-fold dependence, since overlapping training sets induce nonzero covariances between fold losses. One promising approach is to express $\mathrm{Var}(K\text{-CV})$ in terms of per-fold variances and average cross-fold correlations, yielding an ``effective number of folds'' that interpolates between the independent-fold regime (variance reduction $\approx 1/K$) and the fully dependent regime (no reduction). Establishing sharp, assumption-transparent bounds of this kind-- potentially using algorithmic stability or related dependence controls-- would further tighten uncertainty quantification while preserving the interpretability of our hold-out size framework.

More broadly, this framework invites researchers across disciplines to think of cross-validation not as a fixed recipe, but as a design decision tuned to their data and scientific goals.

\paragraph*{Practical recommendations.}
Based on our theoretical and empirical results, we offer the following guidance for applied researchers:

\begin{enumerate}
    \item \textit{Low-noise settings} (controlled experiments, high-precision measurements): 
    The evaluation-uncertainty term is typically small, so it is often reasonable to use a larger $K=N/m$ (e.g., 10--20 folds), prioritizing {training} data to reduce bias.

    \item \textit{High-noise settings} (surveys, behavioral data, many social-science applications): 
    Evaluation uncertainty can dominate, so smaller $K$ (e.g., 2--5 folds) can be preferable to stabilize model evaluation, even at the cost of reduced training size.

    \item \textit{Complex models} (random forests, neural networks, ensembles): 
    These models can be data-hungry, so a larger $K$ (smaller hold-out) is often beneficial when noise is moderate; however, under high noise, the optimal $K$ can shift downward because uncertainty in evaluation dominates.

    \item \textit{Simple models} (linear/logistic regression): 
    These models typically learn effectively from smaller training sets, so the optimal $K$ is often more sensitive to the assumed noise level and the desired precision of model evaluation.

    \item \textit{When uncertain}: 
    Compute the Pareto frontier across a plausible range of $\sigma^2$. If the recommended $K$ is stable across $\sigma^2$ (e.g., $K=5$--$10$ over $\sigma^2 \in [0.1,1]$), report this robustness range. If it varies substantially, report results under multiple splits and interpret conclusions as sensitive to noise assumptions.

    \item \textit{For model comparison}: 
    Use the same $\sigma^2$ assumption when comparing models on the same dataset, but allow that the optimal $K$ may differ across models. When conclusions change across $K$, the Pareto frontier helps diagnose whether the disagreement reflects genuine performance differences or sensitivity to evaluation uncertainty.
\end{enumerate}

As a diagnostic tool, reverse-engineering the implied $\sigma^2$ under conventional splits (Table~3) can reveal whether standard practice aligns with domain knowledge about data quality. Because $\sigma^2$ is scale-dependent, this diagnostic is most interpretable after standardizing the outcome (or otherwise fixing its measurement scale). Implausible implied values under that scale (e.g., unusually large noise for a well-controlled setting) suggest that conventional $K$ is inappropriate for the problem at hand.

\paragraph{Practical implementation for $K$-fold CV.}
Our variance bound is derived for the evaluation loss on a {single} hold-out set of size $m$, and is therefore the appropriate conservative choice when a practitioner uses a single train/test split or wishes to retain a worst-case guarantee at a given hold-out size. In many applications, however, researchers report the {$K$-fold averaged} cross-validation estimate, which aggregates $K$ fold losses and typically has substantially smaller variance than a single hold-out evaluation. In this setting, the single-split bound remains valid but can be conservative. For practitioners who wish to approximate the uncertainty of the reported $K$-fold average more closely, we  consider a simple CLT-style rescaling of the bound as a heuristic ``practical'' variance proxy, and evaluate its accuracy empirically. Across all tested noise regimes and hold-out sizes, this practical proxy closely tracks the Monte Carlo variance of the $K$-fold estimator well for $K\geq4$ (Appendix Figure~\ref{fig:kfold}). While this heuristic does not carry the same finite-sample guarantee as the theoretical bound, it provides a useful operational approximation.

\bibliographystyle{elsarticle-harv}
\bibliography{pnas-sample} 

\clearpage
\newpage
\section*{Appendix}

\subsection*{Additional Simulation Results}

This appendix provides supplementary numerical results illustrating (i) the
smoothness and robustness of the cross-validation loss curve as a function of
the test-set size $m$, (ii) the empirical performance of the theoretical
hold-out variance bound, and (iii) the behavior of several variance
estimators in $K$-fold cross-validation. All simulations use linear regression
with $N=1000$ observations, $p=10$ predictors, and a fixed
data-generating process (A.4) unless otherwise noted.

\subsubsection*{A.1 \; Shape of the CV loss as a function of test size $m$}

\begin{figure}[h!]
    \centering
    \includegraphics[width=0.45\textwidth]{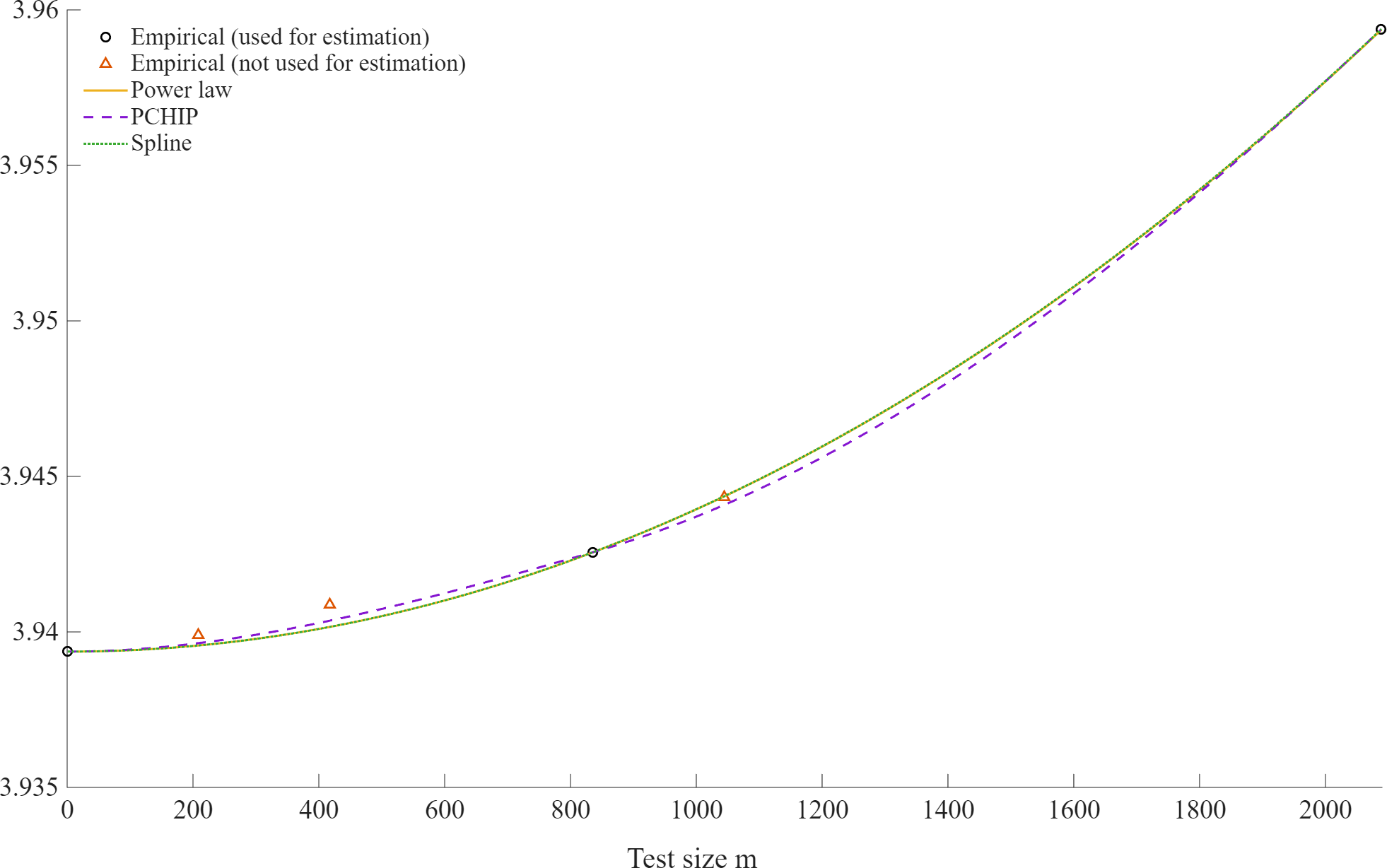}
    \caption{Empirical CV loss and fitted curves (power-law, PCHIP, spline).
    Circles denote points used for fitting; triangles denote withheld points.}
    \label{fig:loss_fits}
\end{figure}

Figure~\ref{fig:loss_fits} plots the empirical cross-validation loss
$\widehat{L}(m)$ as a function of the test-set size $m$ for a grid of values for the Abalone dataset.
Three smooth approximations are:  a power-law fit (as is done in the paper), a shape-preserving PCHIP (shape-preserving piecewise cubic Hermite interpolating polynomial) interpolation, and a cubic spline.

Although only a subset of the empirical points (circles) is used for the
fitting, all three methods produce nearly identical curves. The unused empirical
points (triangles) lie directly on, or near, the fitted curves. The loss is 
smooth and exhibits a slowly increasing convex trend in $m$, consistent with the
fact that larger $m$ implies a smaller training set and hence a modest increase
in prediction error. This motivates using smooth parametric or nonparametric
approximations for $\mathbb{E}[L(m)]$ in theoretical expressions.

\subsubsection*{A.2 \; Hold-out CV: empirical variance vs.\ theoretical bounds}

\begin{figure}[h!]
    \centering
    \includegraphics[width=0.45\textwidth]{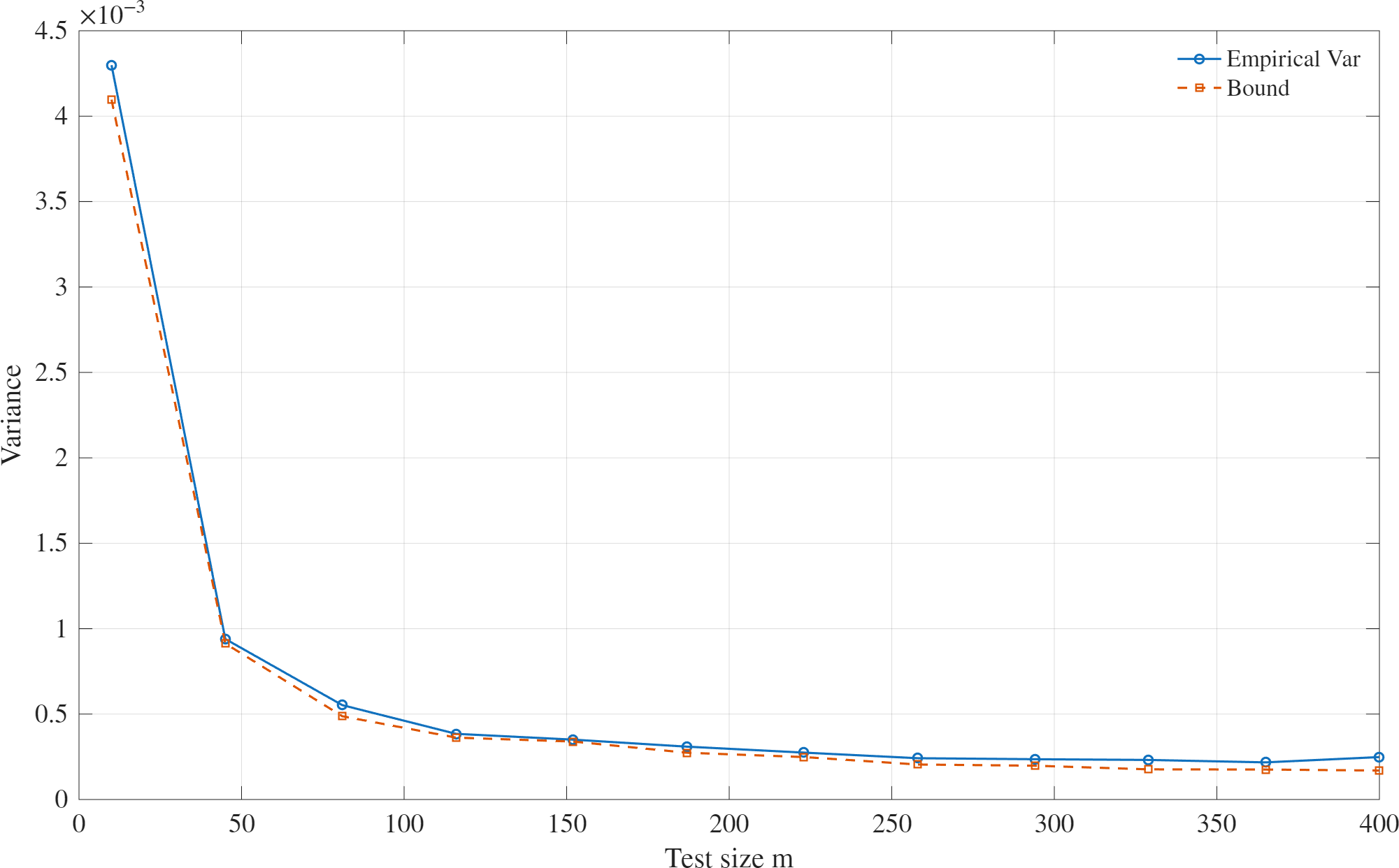}
    \caption{CV variance (Gaussian noise): empirical vs.\ bound.}
    \label{fig:gaussian}
\end{figure}

\begin{figure}[h!]
    \centering
    \includegraphics[width=0.45\textwidth]{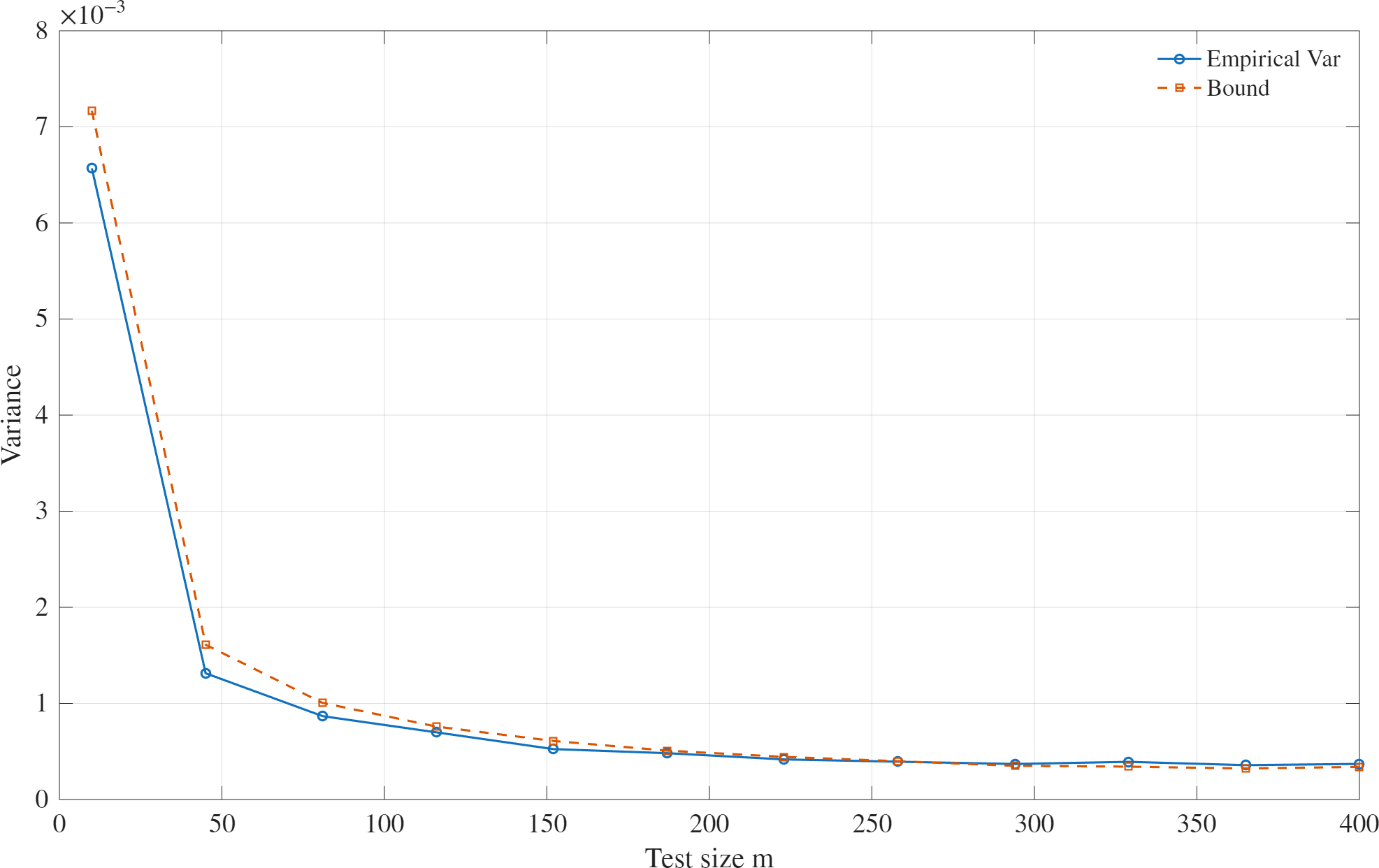}
    \caption{CV variance (heteroskedastic noise): empirical vs.\ bound.}
    \label{fig:hetero}
\end{figure}

\begin{figure}[h!]
    \centering
    \includegraphics[width=0.45\textwidth]{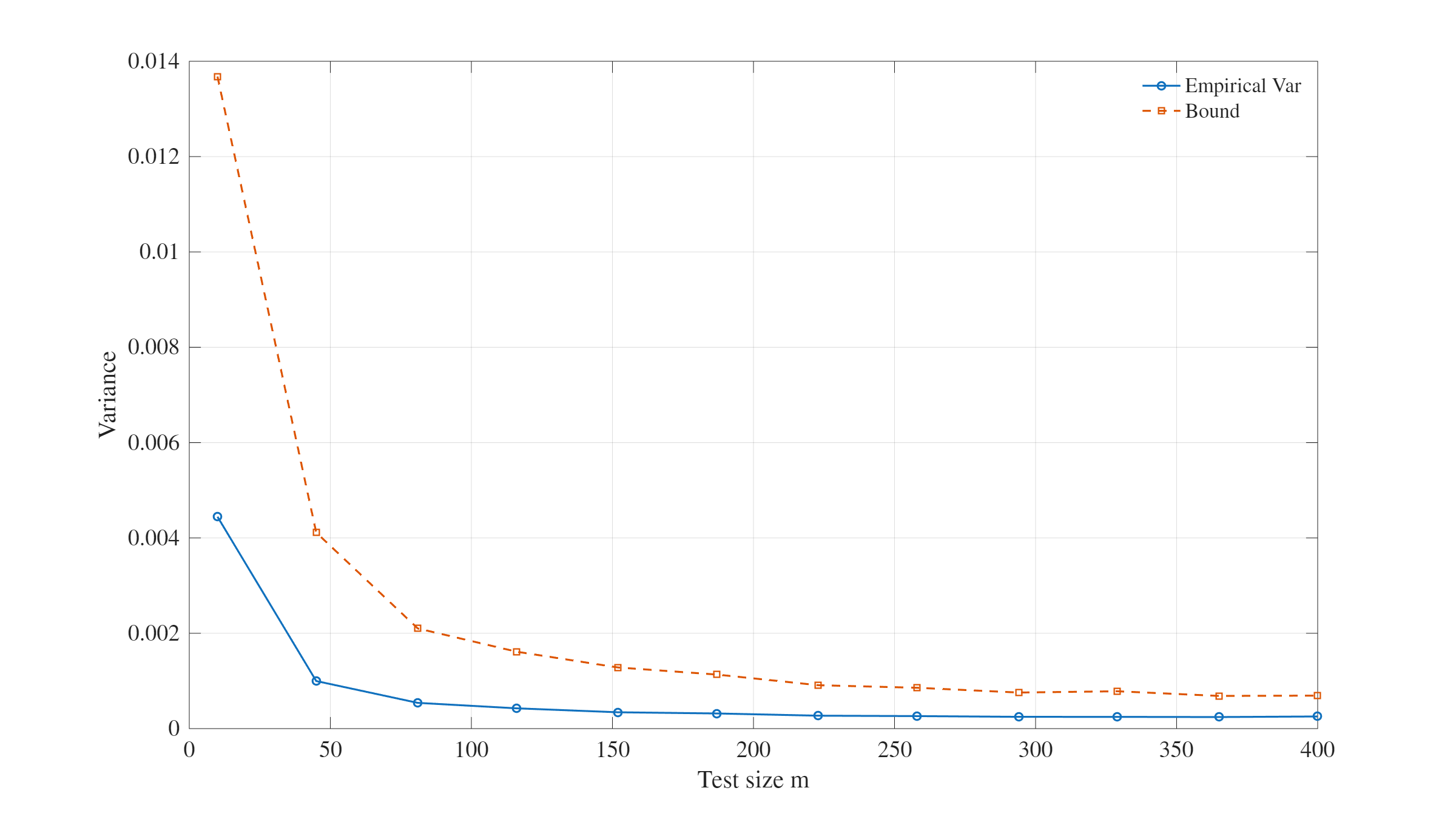}
    \caption{CV variance (skewed noise): empirical vs.\ bound.}
    \label{fig:skewed}
\end{figure}

We now examine the behavior of the theoretical bound derived in the main
paper for the hold-out cross-validation variance,
\[
    \mathrm{Var}\bigl(L_{\mathrm{simple}}(m)\bigr)
    \;\le\;
        C \,\frac{\sigma^2}{m} \, \mathbb{E}[L(m)].
\]
Figures~\ref{fig:gaussian}--\ref{fig:skewed} show the empirical variance of the
practical loss over repeated data draws, together with the bound
$C\sigma^2 \mathbb{E}[L(m)]/m$ for three noise models:
homoskedastic Gaussian noise,
heteroskedastic Gaussian noise, and centered Gamma (skewed) noise.

Several robust patterns appear across all noise regimes:
\begin{enumerate}
\item The empirical variance decreases monotonically in $m$.  
      This follows directly from the $1/m$ structure of the practical loss:
      averaging over more test points reduces randomness from the test set.
\item The theoretical bound is almost exact for the symmetric cases, and bounded by a factor
      of at most $3$ for the skewed case.  
      This is expected and confirms our theoretical results.
\item The gap between the empirical variance and the bound narrows as $m$
      increases, because both quantities scale approximately as $1/m$.
\end{enumerate}

\subsubsection*{A.3 \; $K$-fold CV: bounds, true variance, and variance estimators}

\begin{figure*}[h!]
    \centering
    \includegraphics[width=0.9\textwidth]{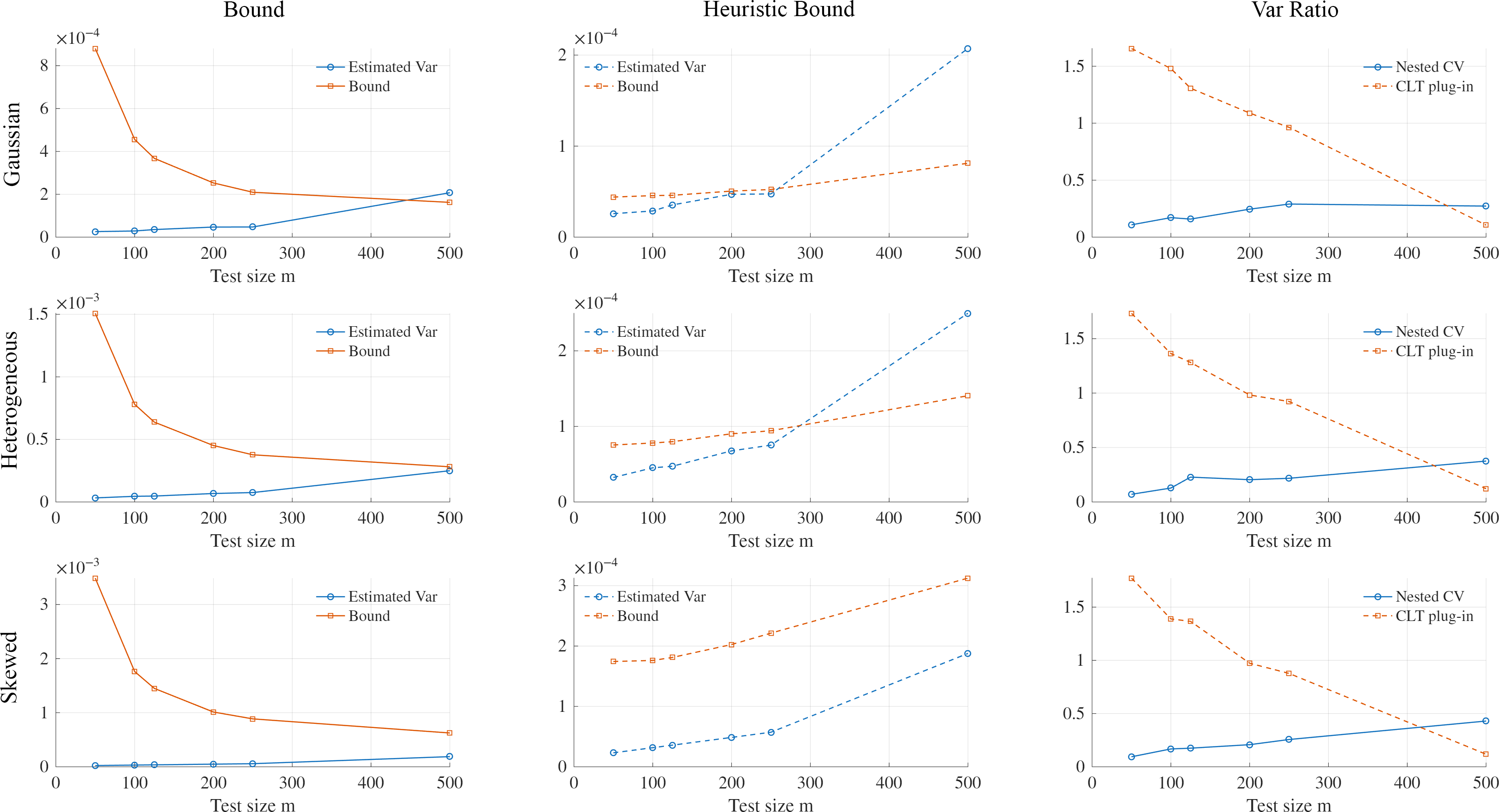}
    \caption{K-fold CV variance comparison:
    bound vs.\ true variance (left column), practical bound vs.\ true variance (middle column)
    and competing estimators vs.\ true variance (right column), for the Gaussian case (top row), heterogeneous case (middle row), and skewed case (bottom row). $C$ denotes the constant on the bound.}
    \label{fig:kfold}
\end{figure*}

Figure~\ref{fig:kfold} presents  how the theoretical bound behaves when cross-validation averages over
$K$ folds.  For each noise model, the left, and middle, panels plot the estimated variance and our bound,
while the right-hand panels plot
\[
    \frac{\widehat{\mathrm{Var}}_{\mathrm{Nested}}}
         {\text{TrueVar}_{K\text{-CV}}}
    \qquad\text{and}\qquad
    \frac{\widehat{\mathrm{Var}}_{\mathrm{CLT}}}
         {\text{TrueVar}_{K\text{-CV}}}.
\]

The findings can be summarized as follows.

\paragraph{Split variance bound is extremely conservative.}
The split constants produce variances with large dependencies with small $m$ that decreases as $m$ increases. This is expected as small $m$ implies large $K$ and thus lower variance due to more folds to test.

\paragraph{CLT-adjusted $K$-fold bounds are surprisingly tight.}
The curves corresponding to $C/(4K)$ and $C/(16K)$ lie very close to~$1$ across
all values of $m$ (except for $m=2$) and for all noise types. This indicates that while the
variance bound is loose, the averaging intrinsic to $K$-fold CV tightens the
variance by a factor of approximately $1/K$, yielding a practically accurate
estimate/bound for the true variance for most relevant $K$.

\paragraph{Variance estimators are strongly unstable.}
Both the Nested-CV estimator and the CLT plug-in estimator substantially
over- and underestimate the true variance.
Practical estimators fail to
capture the sensitivity of the fitted model to the randomness in the training
set; the dominant component of cross-validation variance.

\subsection*{A.4 \; Data Generating Processes for All Simulations}

This section summarizes the data generating processes used in all appendix
figures.  The same basic linear model is used throughout, with differences only
in (i) the noise distribution, and (ii) the manner in which cross-validation is
performed (split vs.\ $K$-fold averaging).

\paragraph{Design and signal.}
All simulations use a fixed design matrix
\[
X \in \mathbb{R}^{N\times p}, \qquad X_{ij}\sim N(0,1),
\]
with $N\in\{800,1000\}$ and $p\in\{10,15\}$ depending on the experiment.  
The true regression function is linear:
\[
f^{\star}(x) = x^{\top}\beta,
\qquad
\beta = (0.1,\,1.2,\,0.8,\,0,\ldots,0)^{\top},
\]
so that only the first three predictors influence the outcome.

\paragraph{Noise distributions.}
Independent errors $\varepsilon\in\mathbb{R}^{N}$ are generated from one of:
\[
\begin{aligned}
&\text{(Gaussian)} && \varepsilon_i\sim N(0,\sigma^{2}), \\
&\text{(Heteroskedastic)} &&
    \varepsilon_i \sim N(0,{\sigma^{2}(1+0.5Z_i)},
    \;\; Z_i\sim U(0,1), \\
&\text{(Skewed / Gamma)} &&
    \varepsilon_i = G_i - k\theta,\;\;
    G_i\sim\mathrm{Gamma}(k,\theta),\ \\
    & && k=5,\ \theta=\sqrt{\sigma^{2}/k}.
\end{aligned}
\]
The observed response is $Y = f^{\star}(X)+\varepsilon$ with $\sigma^{2}=1$.

\paragraph{Model fitting.}
In all settings, the estimator is ordinary least squares trained on the relevant
training subset:
\[
\widehat{\beta} = \arg\min_{b}\|Y_{\text{train}} - X_{\text{train}}b\|^{\,2},
\qquad
\widehat{f}(x)=x^{\top}\widehat{\beta}.
\]

\subsubsection*{A.4.1 \; Split CV (Figures~\ref{fig:gaussian}--\ref{fig:skewed})}

For each test-set size $m\in\{10,\dots,400\}$, we perform $R=1000$ repetitions:
\begin{enumerate}
    \item Draw new noise $\varepsilon$ and set $Y=f^{\star}(X)+\varepsilon$.
    \item Select a random test subset $S\subset\{1,\ldots,N\}$ of size $m$;
          train on $S^{c}$ and test on $S$.
    \item Compute  the unbiased estimate of the pure loss   
        \[
        L_{\mathrm{pure}}(m)
          = \frac1m\sum_{j\in S}(\widehat{f}(X_j)-Y_j)^{2} - \frac1m\sum_{j\in S}\varepsilon_j^{2}.
        \]
    \item Record  
          $\mathrm{Var}(L_{\mathrm{pure}})$ and
          $\mathbb{E}[L_{\mathrm{pure}}]$.
    \item Compute the paper’s single-split bound  
        \[
        \mathrm{Bound}(m)
            = \frac{4\sigma^{2}}{m}\,\mathbb{E}[L_{\mathrm{pure}}(m)].
        \]
\end{enumerate}
These quantities produce the split variance curves and the tightness
comparisons in Figures~\ref{fig:gaussian}--\ref{fig:skewed}.

\subsubsection*{A.4.2 \; $K$-Fold CV (Figure~\ref{fig:kfold})}

For the $K$-fold diagrams, we fix $(N,p)$ and the design $(X,f^{\star})$ and
examine variance as a function of $K$, equivalently $m=N/K$.

For each $K$, we evaluate:
\begin{itemize}
    \item \textbf{True MC variance of $K$-fold CV:}  
        Repeatedly draw new noise $\varepsilon$, form $Y$, generate a fresh
        $K$-fold partition, compute
        \[
        \widehat{L}_{K}
            = \frac1K\sum_{k=1}^{K}
              \frac{1}{m_k}\sum_{j\in S_k}
                (\widehat{f}_{-k}(X_j)-f^{\star}(X_j))^{2},
        \]
        and estimate $\mathrm{Var}(\widehat{L}_K)$ over repetitions.

    \item \textbf{Nested-CV variance (anti-conservative):}  
        Fix a single $Y$, redraw multiple $K$-fold partitions, and compute the
        variance across partitions.

    \item \textbf{CLT plug-in estimator:}  
        Estimate the fold-to-fold variance of the per-fold losses $L_k$ and use
        $\widehat{\mathrm{Var}}_{\mathrm{CLT}}=\widehat{\mathrm{Var}}(L_k)/K$.

    \item \textbf{Bound scaling for $K$-fold CV:}  
        Apply the split bound to each fold and combine via a CLT-style
        factor:
        \[
        \frac{4\sigma^{2}}{m}\mathbb{E}[L(m)],
        \quad
        \frac{4\sigma^{2}}{mK},\quad 
        \frac{16\sigma^{2}}{mK},
        \]
        corresponding to the $C=4$, $C=16$, and CLT-adjusted versions appearing
        in Figure~\ref{fig:kfold}.
\end{itemize}

The resulting comparisons show that:
(i) the raw bound is intentionally conservative,
(ii) the CLT-scaled bound closely tracks the true $K$-fold variance across all
$m$ (except $K\geq4$) and all noise regimes, and
(iii) practical variance estimators severely underestimate variance, especially
for small $m$.

\subsection*{A.5 Proof of Theorem~\ref{thm:var}}

\begin{proof}
In the following, expectations are taken conditional on the training set and test features, treating the trained model predictions as fixed constants.

First, we introduce the following theorem:
\begin{thm}
(Theorem 1 of \cite{Whittle_60}) Let the random variables $z_{1},z_{2},\ldots,z_{n}$
be statistically independent, and each assume the values $\pm1$ with
respective probabilities $\frac{1}{2}$ . Then, for any set of real
coefficients $b_{j}$, $s\geq2$,
\begin{align}
\mathbb{E}\left[\left|\sum_{j}^{n}b_{j}z_{j}\right|^{s}\right] & \leq C\left(s,n\right)\left(\sum_{j}b_{j}^{2}\right)^{s/2}\label{eq:Whittle_Th.1}
\end{align}
where $C\left(s.n\right)=\frac{1}{2^{n}}\frac{1}{n^{s/2}}\sum_{k=0}^{n}\left(\begin{array}{c}
n\\
k
\end{array}\right)\left|n-2k\right|^{s}$. The equality sign holds in \eqref{eq:Whittle_Th.1} if $s=2$ or
if all the $b_{j}$ have equal modulus, but not otherwise. 
\end{thm}

From direct calculation, we have
\begin{alignat*}{1}
 & \left(\frac{1}{m}\left\Vert \hat{\boldsymbol{\mu}}_{k}-\boldsymbol{y}_{k}\right\Vert _{\mathbb{R}^{m}}^{2}-\frac{1}{m}\left\Vert \boldsymbol{\varepsilon}\right\Vert _{\mathbb{R}^{m}}^{2}\vphantom{\mathbb{E}_{Y}\left[\left.\frac{1}{m}\left\Vert \hat{\boldsymbol{\mu}}_{k}-\boldsymbol{y}_{k}\right\Vert _{\mathbb{R}^{m}}^{2}-\frac{1}{m}\left\Vert \boldsymbol{\varepsilon}\right\Vert _{\mathbb{R}^{m}}^{2}\right|\boldsymbol{x}\right]}\right.\\
 & \left.\vphantom{\left(\frac{1}{m}\left\Vert \hat{\boldsymbol{\mu}}_{k}-\boldsymbol{y}_{k}\right\Vert _{\mathbb{R}^{m}}^{2}\right.}-\mathbb{E}_{Y}\left[\left.\frac{1}{m}\left\Vert \hat{\boldsymbol{\mu}}_{k}-\boldsymbol{y}_{k}\right\Vert _{\mathbb{R}^{m}}^{2}-\frac{1}{m}\left\Vert \boldsymbol{\varepsilon}\right\Vert _{\mathbb{R}^{m}}^{2}\right|\boldsymbol{x}\right]\right)^{2}\\
= & \left(\frac{1}{m}2\boldsymbol{\varepsilon}^{\top}\left(\hat{\boldsymbol{\mu}}_{k}-\boldsymbol{\mu}_{k}\right)\right)^{2}
\end{alignat*}
We now aim to obtain, 
$\mathbb{E}\left[\left.\left(\frac{1}{m}2\boldsymbol{\varepsilon}_{k}^{\top}\left(\hat{\boldsymbol{\mu}}_{k}-\boldsymbol{\mu}_{k}\right)\right)^{2}\right|\boldsymbol{x}\right]$.
Within the $\boldsymbol{x}$-given expectation operator, the $\left(\hat{\boldsymbol{\mu}}_{k}-\boldsymbol{\mu}_{k}\right)$ term can be treated as a constant.
Let
$\mathbb{E}_{\pm}\left[\cdot\right]$ denote an expectation with respect
to the sign of the $\varepsilon$. Since $\varepsilon$ is symmetric, we have
\begin{align*}
\mathbb{E}&\left[\left.\left|\sum_{j=1}^{m}\varepsilon_{j}\left(\hat{\mu}_{j}-\mu_{j}\right)\right|^{2}\right|\boldsymbol{x}\right] \\
& =\mathbb{E}\left[\left.\mathbb{E}_{\pm}\left[\left.\left|\sum_{j=1}^{m}\textrm{sgn}\left(\varepsilon_{j}\right)\left|\varepsilon_{j}\right|\left(\hat{\mu}_{j}-\mu_{j}\right)\right|^{2}\right|\boldsymbol{x}\right]\right|\boldsymbol{x}\right].
\end{align*}
and from \eqref{eq:Whittle_Th.1}, we have
\begin{align*}
 & \mathbb{E}\left[\left.\mathbb{E}_{\pm}\left[\left.\left|\sum_{j=1}^{m}\textrm{sgn}\left(\varepsilon_{j}\right)\left|\varepsilon_{j}\right|\left(\hat{\mu}_{j}-\mu_{j}\right)\right|^{2}\right|\boldsymbol{x}\right]\right|\boldsymbol{x}\right]\\
= & \mathbb{E}\left[\left.C\left(2,m\right)\left|\sum_{j=1}^{m}\left|\varepsilon_{j}\right|^{2}\left(\hat{\mu}_{j}-\mu_{j}\right)^{2}\right|\right|\boldsymbol{x}\right]\\
= & C\left(2,m\right)\left(\sum_{j\in C_{k}}\sigma_{j}^{2}\left(\hat{f}_{-k}(x_{j})-f(x_{j})\right)^{2}\right).
\end{align*}
Further, since $C\left(2,m\right)$ is monotone increasing regarding $m$, we have
\[
C\left(2,m\right)=1.
\]
\end{proof}

\subsection*{A.6 Proof of Theorem~\ref{thm:var2}}

\begin{proof}
When $\varepsilon$ is not symmetric, let $\tilde{\varepsilon}_{j}$ be the copy of $\varepsilon_{j}$ (sampled from the same distribution, independently from $\varepsilon_{j}$).
Then from the symmetry of $\varepsilon_{j}-\tilde{\varepsilon}_{j}$, we have
\begin{align*}
\mathbb{E}\left[\left.\left|\sum_{j=1}^{m}\varepsilon_{j}\left(\hat{\mu}_{j}-\mu_{j}\right)\right|^{2}\right|\boldsymbol{x}\right] & \leq\mathbb{E}\left[\left.\left|\sum_{j=1}^{m}\left(\varepsilon_{j}-\tilde{\varepsilon}_{j}\right)\left(\hat{\mu}_{j}-\mu_{j}\right)\right|^{2}\right|\boldsymbol{x}\right].
\end{align*}
As with the case when $\varepsilon_{j}$ is symmetric, from \eqref{eq:Whittle_Th.1}, we have
\begin{align*}
 & \mathbb{E}\left[\left.\mathbb{E}_{\pm}\left[\left.\left|\sum_{j=1}^{m}\textrm{sgn}\left(\varepsilon_{j}-\tilde{\varepsilon}_{j}\right)\left|\varepsilon_{j}-\tilde{\varepsilon}_{j}\right|\left(\hat{\mu}_{j}-\mu_{j}\right)\right|^{2}\right|\boldsymbol{x}\right]\right|\boldsymbol{x}\right]\\
= & \mathbb{E}\left[\left.C\left(2,m\right)\left|\sum_{j=1}^{m}\left|\varepsilon_{j}-\tilde{\varepsilon}_{j}\right|^{2}\left(\hat{\mu}_{j}-\mu_{j}\right)^{2}\right|\right|\boldsymbol{x}\right].
\end{align*}
Further, from the Minkowski inequality,
\[
\mathbb{E}\left[\left|\varepsilon_{j}-\tilde{\varepsilon}_{j}\right|^{2}\right]\underset{\textrm{Minkowski}}{\leq}\left(\mathbb{E}\left[\left|\varepsilon_{j}\right|^{2}\right]^{1/2}+\mathbb{E}\left[\left|\tilde{\varepsilon}_{j}\right|^{2}\right]^{1/2}\right)^{2}=\left(2\sigma_{j}\right)^{2}.
\]
Therefore,
\begin{align*}
 & \mathbb{E}\left[\left.C\left(2,m\right)\left|\sum_{j=1}^{m}\left|\varepsilon_{j}-\tilde{\varepsilon}_{j}\right|^{2}\left(\hat{\mu}_{j}-\mu_{j}\right)^{2}\right|\right|\boldsymbol{x}\right]\\
\leq & 4C\left(2,m\right)\mathbb{E}\left[\left.\left|\sum_{j=1}^{m}\sigma_{j}^{2}\left(\hat{\mu}_{j}-\mu_{j}\right)^{2}\right|\right|\boldsymbol{x}\right].
\end{align*}
\end{proof}

\subsection*{A.7 Fold averaging and variance}
Let $L_1,\dots,L_K$ denote the losses evaluated on $K$ folds of equal test size $m$, and let
\[
\bar L = \frac{1}{K}\sum_{k=1}^K L_k
\]
denote the $K$-fold cross-validation estimator. Then
\[
\mathrm{Var}(\bar L)
= \frac{1}{K^2}\sum_{k=1}^K \mathrm{Var}(L_k)
+ \frac{2}{K^2}\sum_{k<\ell} \mathrm{Cov}(L_k,L_\ell).
\]
By the Cauchy--Schwarz inequality, $|\mathrm{Cov}(L_k,L_\ell)| \le \sqrt{\mathrm{Var}(L_k)\mathrm{Var}(L_\ell)}$, and hence
\[
\mathrm{Var}(\bar L)
\le \frac{1}{K^2}\Big( K + K(K-1) \Big)\max_k \mathrm{Var}(L_k)
= \max_k \mathrm{Var}(L_k).
\]
Therefore, any upper bound on the variance of a single fold loss also upper-bounds the variance of the $K$-fold averaged estimator.
\end{document}